\documentclass[twocolumn,prb,color,superscriptaddress,psfig,showpacs,amsmath,amssymb,fleqn]{revtex4-1}

\usepackage{amsmath}
\usepackage{amsfonts}
\usepackage{amssymb}
\usepackage{amsthm}
\usepackage{epsf}

\usepackage{natbib}
\setcitestyle{square,numbers}

\usepackage{tabularx} 
\usepackage{graphicx} 
\usepackage{epstopdf}
\usepackage{epsf}
\usepackage{dcolumn}
\usepackage{bm}
\usepackage{color}

\newcommand{\ICVO}{InCu$_{2/3}$V$_{1/3}$O$_3$}

\usepackage{multirow}

\usepackage[normalem]{ulem}

\usepackage{color,soul}
\setul{0.5ex}{0.5ex}
\setulcolor{red}

\makeatletter
\newcommand\cwave[1][red]{\bgroup \markoverwith{\lower5\p@\hbox{\sixly \textcolor{#1}{\char58}}}\ULon}
\font\sixly=lasyb10 
\makeatother

\definecolor{mygray}{cmyk}{0, 0, 0, 0.3}
\definecolor{Ora}{cmyk}{0, 0.6, 0.8, 0.4}
\definecolor{ojcolor}{cmyk}{0.63, 0.33, 0, 0.26}

\newcommand{\out}[1]{{\color{mygray}\sout{#1}}}
\newcommand{\vk}[1]{{\color{blue}{#1}}}

\begin{document}

\title{Ground state and low-temperature magnetism of the quasi-two-dimensional honeycomb compound \ICVO}

\author{M. Iakovleva}
\affiliation{Leibniz Institute for Solid State and Materials Research IFW Dresden, 01171 Dresden, Germany}
\affiliation{Institute for Solid State and Materials Physics, TU Dresden, 01069 Dresden, Germany}
\affiliation{Zavoisky Physical-Technical Institute, FRC Kazan
	Scientific Center of RAS, 420029 Kazan, Russia}

\author{O. Janson}
\affiliation{Leibniz Institute for Solid State and Materials Research IFW Dresden, 01171 Dresden, Germany}

\author{H.-J. Grafe}
\affiliation{Leibniz Institute for Solid State and Materials Research IFW Dresden, 01171 Dresden, Germany}

\author{A. P. Dioguardi}
\affiliation{Leibniz Institute for Solid State and Materials Research IFW Dresden, 01171 Dresden, Germany}

\author{H. Maeter}
\affiliation{Institute for Solid State and Materials Physics, TU Dresden, 01069 Dresden, Germany}

\author{N. Yeche}
\affiliation{Institute for Solid State and Materials Physics, TU Dresden, 01069 Dresden, Germany}

\author{H.-H. Klauss}
\affiliation{Institute for Solid State and Materials Physics, TU Dresden, 01069 Dresden, Germany}

\author{G. Pascua}
\affiliation{ Labor f\"{u}r Myonenspinspektroskopie, Paul Scherrer Institut, CH-5232 Villigen PSI, Switzerland}

\author{H. Luetkens}
\affiliation{ Labor f\"{u}r Myonenspinspektroskopie, Paul Scherrer Institut, CH-5232 Villigen PSI, Switzerland}

\author{A. \ M\"{o}ller}
\affiliation{JGU Mainz, 55122 Mainz, Germany}

\author{B.\ B\"{u}chner}
\affiliation{Leibniz Institute for Solid State and Materials Research IFW Dresden, 01171 Dresden, Germany} 
\affiliation{Institute for Solid State and Materials Physics, TU Dresden, 01069 Dresden, Germany}
\affiliation{W\"{u}rzburg–Dresden Cluster of Excellence ct.qmat}

\author{V. Kataev}
\affiliation{Leibniz Institute for Solid State and Materials Research IFW Dresden, 01171 Dresden, Germany}

\author{E. Vavilova}
\affiliation{Zavoisky Physical-Technical Institute, FRC Kazan
	Scientific Center of RAS, 420029 Kazan, Russia}

\date{November 11, 2019}

\begin{abstract}

We report a combined $^{115}$In NQR, $^{51}$V NMR and $\mu$SR spectroscopic
study of the low-temperature magnetic properties of \ICVO, a quasi-two
dimensional (2D) compound comprising in the spin sector a honeycomb lattice of
antiferromagnetically coupled spins $S=1/2$ associated with Cu$^{2+}$ ions.
Despite substantial experimental and theoretical efforts, the ground state of
this material was has not been ultimately identified. In particular, two
characteristic temperatures of about $\sim 40$\,K and $\sim 20$\,K manifesting
themselves as anomalies in different magnetic measurements are discussed controversially.
A combined analysis of the experimental data complemented with theoretical
calculations of exchange constants enabled us to identify below 39\,K an
``intermediate'' quasi-2D static spin state. This spin state is characterized by a staggered
magnetization with a temperature evolution that
agrees with the
predictions for the 2D XY model. We observe that this state gradually
transforms at 15\,K into a fully developed 3D antiferromagnetic N\'eel state.
We ascribe such an extended quasi-2D static regime to an effective magnetic
decoupling of the honeycomb planes due to a strong frustration of the
interlayer exchange interactions which inhibits long-range spin-spin
correlations across the planes.  Interestingly, we find indications of the
topological Berezinsky-Kosterlitz-Thouless transition in the quasi-2D static
state of the honeycomb spin-1/2 planes of \ICVO.

\end{abstract}

\maketitle
\section{Introduction}\label{intro}

Dimensionality of the spin system and the involved spin degrees of freedom  play an important role in defining magnetic properties of a material  in particular regarding its ground state and magnetic excitations. Long-range magnetic order (LRO) which is a typical ground state of 3-dimensional (3D) magnets  can be 
strongly suppressed in systems with frustrated exchange interactions. Reduction of the spatial dimensionalilty of magnetic couplings may be a further cause of the suppression of LRO. For instance, according to the Mermin-Wagner's theorem, LRO is impossible even at zero temperature in a 1D spin-1/2 isotropic Heisenberg chain with the coordination number $z = 2$ due to strong quantum fluctuations, although it can be stabilized on  a 2D square lattice  with $z = 4 $ at $T=0$\,K \cite{Mermin1966}. The spin-1/2 Heisenberg honeycomb lattice  has a particular position  in this respect. Like the square lattice, it is not frustrated for the antiferromagnetic (AFM) nearest-neighbor interaction.  However, quantum fluctuations are enhanced due to the low coordination number $z=3$. Still they remain  less pronounced than in the 1D case and, hence, do not destroy LRO at $T=0$. 

In real 3D transition metal (TM) compounds where magnetic TM ions bonded via ligands form 2D hexagonal layers, such honeycomb spin planes may deviate from the ideal Heisenberg regime. The  anisotropy of the in-plane magnetic exchange as well as a finite \vk{interlayer} coupling can weaken  
quantum fluctuations. As a result,  
long range magnetic order can be stabilized at finite temperatures \cite{PhysRevB.92.144401,PhysRevB.96.024417, PhysRevB.95.094424,PhysRevB.94.214416,PhysRevB.78.184410}. 
Depending on the particular details, such as frustration of the next-neighboring in-plane exchange interaction and/or of the  coupling  between the planes, different types of magnetic structures can be realized \cite{PhysRevB.86.144404, PhysRevB.81.214419, 0953-8984-23-22-226006, 0953-8984-24-23-236002, PhysRevB.95.024401, PhysRevB.89.214413}. If frustration is sufficiently strong the LRO can be suppressed completely \cite{ja901922p, PhysRevLett.105.187201}.

One particularly interesting aspect of this class of materials is that the spin-orbit coupling and the low-symmetry ligand field can give rise to a strong easy-plane magnetic anisotropy of the honeycomb planes made of TM ions which then can be described by the $XY$ model. There, unlike in the Heisenberg model, only two components of the spin $S_{\rm x}$ and $S_{\rm y}$ are interacting on the honeycomb spin plane.
This  can lead to complex magnetic behavior
beyond conventional LRO and associated classical spin-wave excitations, namely the
 formation of the spin vortex gas and the Berezinsky-Kosterlitz-Thouless (BKT) transition to the topologically ordered state of vortex-antivortex pairs predicted in the 2D {\it{XY}} model \cite{ Berezinskii, 0022-3719-6-7-010, 0022-3719-5-11-002, 0022-3719-7-6-005}.

Only a few  antiferromagnets are known so far, where the signatures of this topological transition were  found mostly by analyzing the critical exponents. In the majority of the cases the comparison with the predictions of the BKT theory is complicated due to the residual interplane couplings which eventually yield 3D AFM N\'eel order such as, e.g.,  in  BaNi$_2$(PO$_4$)$_2$ \cite{Regnault1990}, BaNi$_2$V$_2$O$_8$ \cite{PhysRevLett.91.137601, PhysRevB.91.214412, PhysRevB.96.214428}, or
MnPS$_3$ \cite{RONNOW2000676}. As a rare exception, the coordination polymer C$_{36}$H$_{48}$Cu$_{2}$F$_{6}$N$_{8}$O$_{12}$S$_{2}$ does not exhibit the N\'eel order but features magnetic excitations consistent with a BKT scenario \cite{Tutsch2014}. 

The title compound of the present paper  \ICVO\, 
features 2D structural order of magnetic Cu$^{2+}$ ($S$=1/2) ions
on a honeycomb lattice with nonmagnetic V ions in the formal oxidation state 5+ complementing the hexagonal layer \cite{PhysRevB.78.024420}. Occasional Cu/V site inversion gives rise to a finite structural in-plane correlation length of approximately 300\,\AA\  \cite{PhysRevB.78.024420}. Besides being a  candidate to host $d$-wave superconductivity upon doping \cite{1742-6596-827-1-012010}, it also appears to be a rare realization of a honeycomb spin lattice with strongly frustrated interlayer coupling. Despite substantial experimental and theoretical interest in this compound its ground state is still not unambiguously identified. Previous magnetization, Electron Spin Resonance (ESR) and Nuclear Magnetic Resonance (NMR) measurements complemented by Quantum Monte Carlo calculations suggest the onset of the AFM ordered state at $T_{\rm N}=38$\,K \cite{Kataev2005} with the fully developed N\'eel-type collinear AFM sublattices below $\sim 20$\,K \cite{PhysRevB.81.060414} that feature anomalous spin dynamics \cite{10.1143/JPSJ.80.023705}. However, according to Ref.~\cite{PhysRevB.78.024420}, specific-heat, thermal expansion and neutron-diffraction experiments
show no evidence for long range magnetic order down to 1.8\,K. 
As reported in Ref.~\cite{PhysRevB.85.085102}, doping of \ICVO\, with Co leads to AFM LRO,  whereas Zn doping results in the suppression of AFM order. Theoretical studies have shown that  fluctuations arising from the interlayer magnetic frustration destroy 3D magnetic LRO in \ICVO\ \cite{PhysRevB.85.085102, 10.1063/1.4977227}. 

Considering the above mentioned controversies of different kinds of experimental data and their interpretation it is appealing to resolve them and to obtain a unified picture of the ground state and low-temperature magnetism of \ICVO. With this aim, in the present work we investigate experimentally and theoretically the ground
state and low-temperature static and dynamic properties of 
\ICVO. In particular, employing $^{115}$In NQR, $^{51}$V NMR and $\mu$SR
spectroscopies we address the occurrence of the above
menitioned two characteristic temperature scales of  $\sim 40$\,K and
$\sim 20$\,K. Their understanding appears to be crucial for the elucidation of
the low-temperature magnetism of this compound. We find that below 39\,K \ICVO\
develops a staggered magnetization. Its $T$-dependence suggests a quasi-2D
static state of the spin system with predominantly in-plane commensurate long-range
spin-spin correlations. Further on,  we find that a full 3D antiferromagnetic order
develops at a much lower temperature of 15\,K.  These experimental findings
are rationalized within a microscopic spin model based on density functional
theory (DFT) calculations. Its key features are: i) the dominance of $J_1$ and
the irrelevance of magnetic exchanges beyond the nearest neighbors within the
honeycomb planes, i.e.\ $J_1\gg\max{\left(|J_2|,|J_3|,...\right)}$, ii) the XXZ
anisotropy of the nearest neighbor exchange accompanied by a vanishingly small
antisymmetric (Dzyaloshinskii-Moriya) anisotropy, and iii) the presence of a
single frustrated antiferromagnetic exchange between the honeycomb planes. We
argue that a significant frustration of the interlayer exchange in \ICVO\
together with a certain degree of structural disorder are responsible for the two
characteristic low-temperature regimes of this compound in compliance with all experimental observations.  Furthermore, by
analyzing our static and dynamic magnetic data we find indications of a BKT
transition in \ICVO\ presumably occurring within an intermediate  quasi-2D
static state of the spin system below 39\,K.

The paper is organized as follows. Experimental details are summarized in Sect.~\ref{methods}. The results of $^{115}$In NQR, $^{51}$V NMR and $\mu$SR experiments and the microscopic model are presented in Sect.~\ref{results} and Sect.~\ref{model}, respectively, and discussed together in Sect.~\ref{Discussion}. The main conclusions are summarized in Sect.~\ref{conclusion}. The Appendix presents some specific details of the analysis of the NQR Hamiltonian.

\section{Experimental details}\label{methods}

The powder sample of \ICVO\ was synthesized and comprehensively characterized as described in Ref.~\cite{PhysRevB.78.024420}.

NQR experiments were performed on the $^{115}$In nuclei (nuclear spin $I=9/2$, gyromagnetic ratio $\gamma=9.32$\,MHz/T, quadrupolar moment $Q=0.81$\,barns) using the Redstone console and Lap NMR portable 
spectrometers from Tecmag. Frequency swept spectra were recorded with the $\pi/2-\tau-\pi$ pulse sequence, the spin echo signal was integrated at each step. Measurements were performed in the temperature range from 5 to 100\,K. The nuclear relaxation rates in the NQR experiment were measured at the maximum of the spectra with the saturation-recovery method. For that the $(9 \times \pi/2)-\tau_d-\pi/2 - \pi$ \vk{pulse} train was used to saturate the nuclear spin system and the echo intensity was measured as a function of the time delay $\tau_d$ between the pulses. The $^{51}$V NMR spin-lattice relaxation rate was measured with the method of stimulated echo.

A muon spin relaxation ($\mu$SR) experiment on a powder sample of \ICVO\
was performed at the GPS beamline at the Paul Scherrer Institute in Villigen, Switzerland,
between 2~K and 60~K in zero field (ZF) and 50~G external magnetic field applied perpendicular to the forward/backward positron detector pair axis. 
The time dependent corrected asymmetry spectra  were analyzed using the MSRFIT software package \cite{Suter12}.

\section{Results}\label{results}

\subsection{$^{115}$In NQR experiments}
\subsubsection{ $^{115}$In NQR spectra}

The choice of the NQR technique has several advantages over the experimental methods previously used for the studies of \ICVO. As other magnetic resonance spectroscopies, it is a local technique that provides  information  on the static and dynamic properties of  a material on the scale of a few interatomic spacings and does not require long range order to probe the spin structure and magnetic exitations which would be a necessary condition, e.g.,  in the case of neutron scattering. Compared to NMR, NQR is a technique where no external magnetic field is needed, therefore it allows one to  probe the bare, unperturbed Hamiltonian of the electron spin system. Finally,  NQR  is very sensitive not only to fluctuating and static local magnetic fields  generated by electron spins but also to the electric field gradient around the nuclei thus making it possible to study  a local charge environment.

Generally, all nuclei with spin $I>$\,1/2 have a non-spherical charge distribution and an electric quadrupole moment $Q$ associated with this charge. The quadrupole moment interacts with the local crystal electric field gradient (EFG) that originates from a non-symmetric  ionic surrounding. The interaction of  $Q$ with the EFG can be expressed by the following Hamiltonian \cite{Abragam1961}:

\begin{eqnarray}
 \mathcal{H_Q} = \frac{e^2qQ}{4I(2I-1)}\big[3\hat I_z^2-\hat I^2+\frac{\eta}{2}\big( \hat I_+^2+\hat I_-^2\big) \big].
\label{NQRham}
\end{eqnarray}
Here, $e$ is the charge of the proton, $\eta$ is the asymmetry parameter defined as $\eta = \frac{V_{xx}-V_{yy}}{V_{zz}}$. The $V_{ii}$ is the matrix element of the EFG tensor in the principal  coordinate system. The quantity $q$ is commonly defined as $eq=V_{zz}$, and $\hat I_z, \hat I_+$ and $ \hat I_-$ are projections of the nuclear spin $I$ on the $z$-quantization axis given by the EFG and perpendicular to it, respectively. 
In \ICVO\, the $^{115}$In nucleus with the  spin $I=9/2$  is surrounded by six oxygen  ions  forming an asymmetric octahedron. Due to this asymmetry the degeneracy of the NQR transitions $\pm1/2\leftrightarrow \pm3/2$, $\pm3/2\leftrightarrow \pm5/2$, $\pm5/2\leftrightarrow \pm7/2$ and $\pm7/2\leftrightarrow \pm9/2$ is lifted and, as a result, four lines  in the $^{115}$In NQR spectrum with different frequencies $\nu_{Q{\rm i}}$  are expected (see  below and the Appendix for details). 

\begin{figure}[h]
\includegraphics[width=0.9\columnwidth]{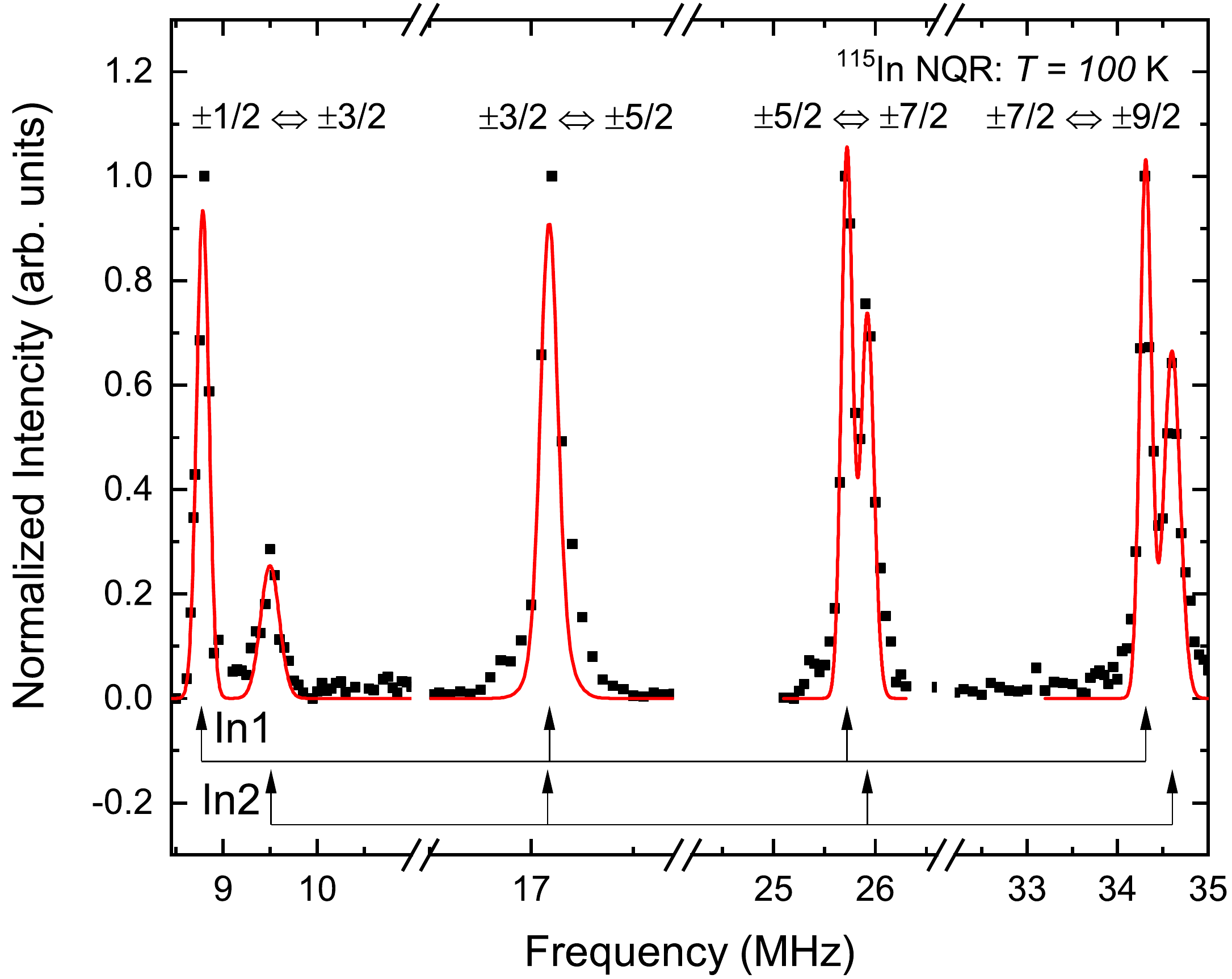}
\caption{ (Color online) High-temperature $^{115}$In NQR spectrum of \ICVO\ (data points). The solid line is a fit to the data according to Hamiltonian (\ref{NQRham}). The positions of the lines corresponding to two non-equivalent In sites in the crystal structure (In1 and In2) are indicated by arrows (see the text for details). For technical reasons all pairs of transitions corresponding to In1 and In2 were measured separately and the intensities of the respective parts of the spectrum were  each normalized to unity.} \label{HTsp}
\end{figure}

\begin{figure}[h]
\includegraphics[width=0.9\columnwidth]{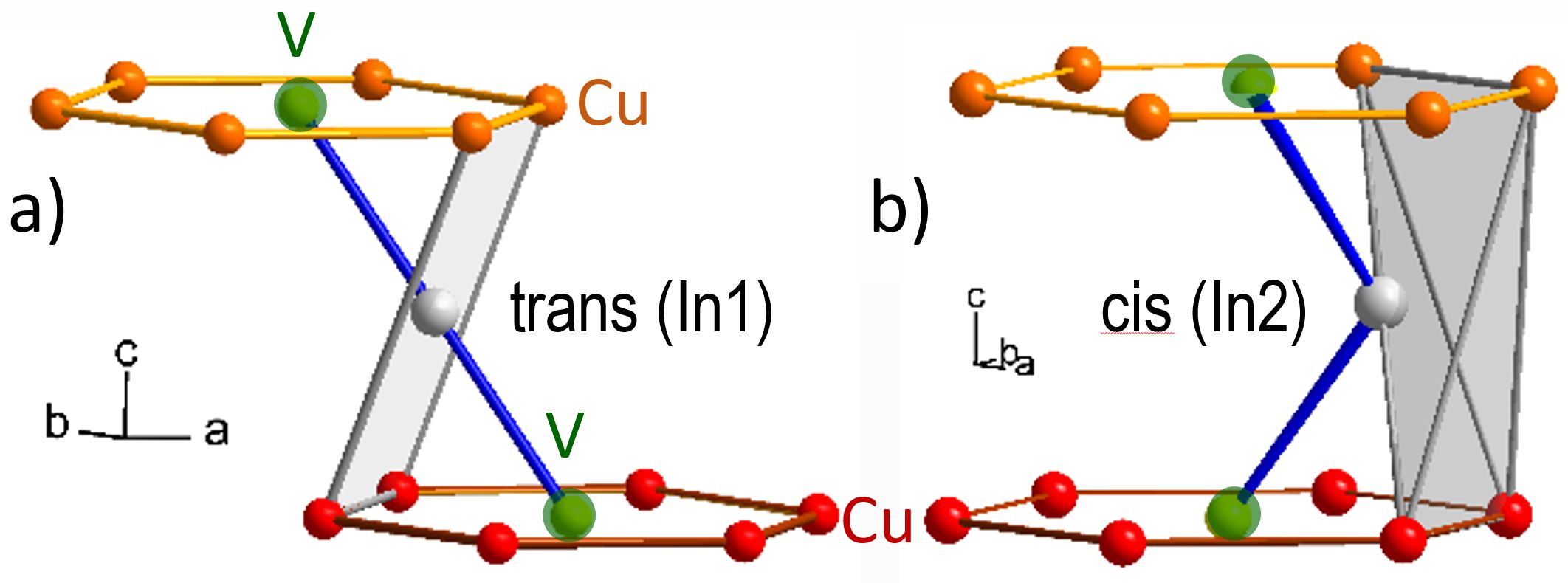}
\caption{ (Color online) Two  crystallographic In1 and In2 positions in \ICVO\ with high (a) and low (b) local  environment symmetry, respectively. }
\label{pos}
\end{figure}

A full  $^{115}$In NQR spectrum of \ICVO\ measured at a high temperature of $T=100$\,K is presented in Fig.~\ref{HTsp}.
Seven  resonance lines can be  resolved instead of the expected four. Such a spectrum can be completely understood and well modeled  if one takes into account two different crystallographic In positions,  In1 and  In2,  \out{arising due to the stacking of ordered honeycomb lattices} \vk{present in the regular crystallographic structure of \ICVO} (Fig.~\ref{pos}). In1 occurs in a centrosymmetric surrounding in a {\it trans}-fashion whereas In2 occurs in a non-centrosymmetric surrounding in a {\it cis}-fashion, thus having slightly different electric field gradients. 
Consequently, for the modeling of the spectra with the aid of Hamiltonian (\ref{NQRham}) we used two set of parameters $\nu_Q$ and $\eta$ which, together with the corresponding transition frequencies, are listed in Table ~\ref{table1}.  Obviously, the smaller $\eta=0.05$  can be assigned to the position with higher symmetry and  the larger $\eta=0.1$  can be assigned to the position with lower symmetry (Fig.~\ref{pos}).  The intensities of the two contributions  are related approximately as 1:1. 
The apparent difference  of the amplitude of the lines in Fig.~\ref{HTsp} is due two their slightly different  widths most likely occuring  because of the different degree of distortion of the oxygen octahedron.    

\begin{table}[h]
	\caption{ Modeling parameters of the $^{115}$In NQR spectrum ($\nu_Q$ and transition frequencies are given in MHz).}
	\begin{ruledtabular}
		\begin{tabular}{c|c c c c c c}
			
			&   &  &\multicolumn{4}{c}{Transition frequencies} \\
			\hline
			
			Site & $\nu_Q$  & $\eta$  &$(\pm\frac{1}{2}, \pm\frac{3}{2})$& ($\pm\frac{3}{2}, \pm\frac{5}{2}$) &($\pm\frac{5}{2}, \pm\frac{7}{2}$) & ($\pm\frac{7}{2},\pm\frac{9}{2}$) \\

			\colrule
			
			In1 & 8.58 & 0.05 & 8.77  & 17.09 &25.72  &34.31  \\
			In2 & 8.66 & 0.1 & 9.51 & 17.08  & 25.92 & 34.60\\
			
		\end{tabular}
	\end{ruledtabular}
	\label{table1}
\end{table}
\begin{figure}[h]
\includegraphics[width=0.9\columnwidth]{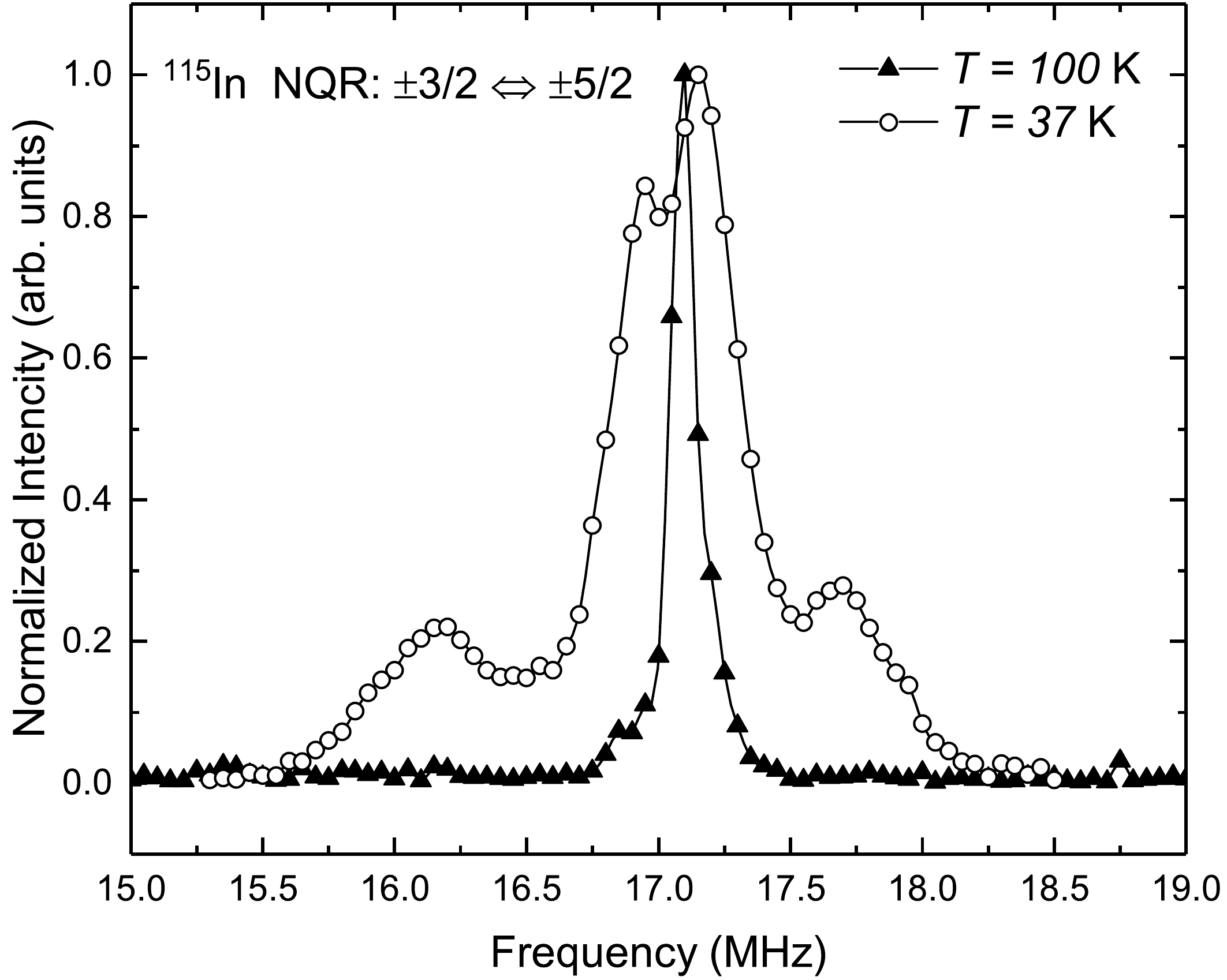}
\caption{$^{115}$In NQR spectra corresponding to the ($\pm\frac{3}{2} \leftrightarrow \pm\frac{5}{2}$) transition (cf. Fig.~\ref{HTsp})  at temperatures of 100\,K (triangles) and 37\,K(circles).} \label{HLsp}
\end{figure}
\begin{figure}[h]
\includegraphics[width=0.9\columnwidth]{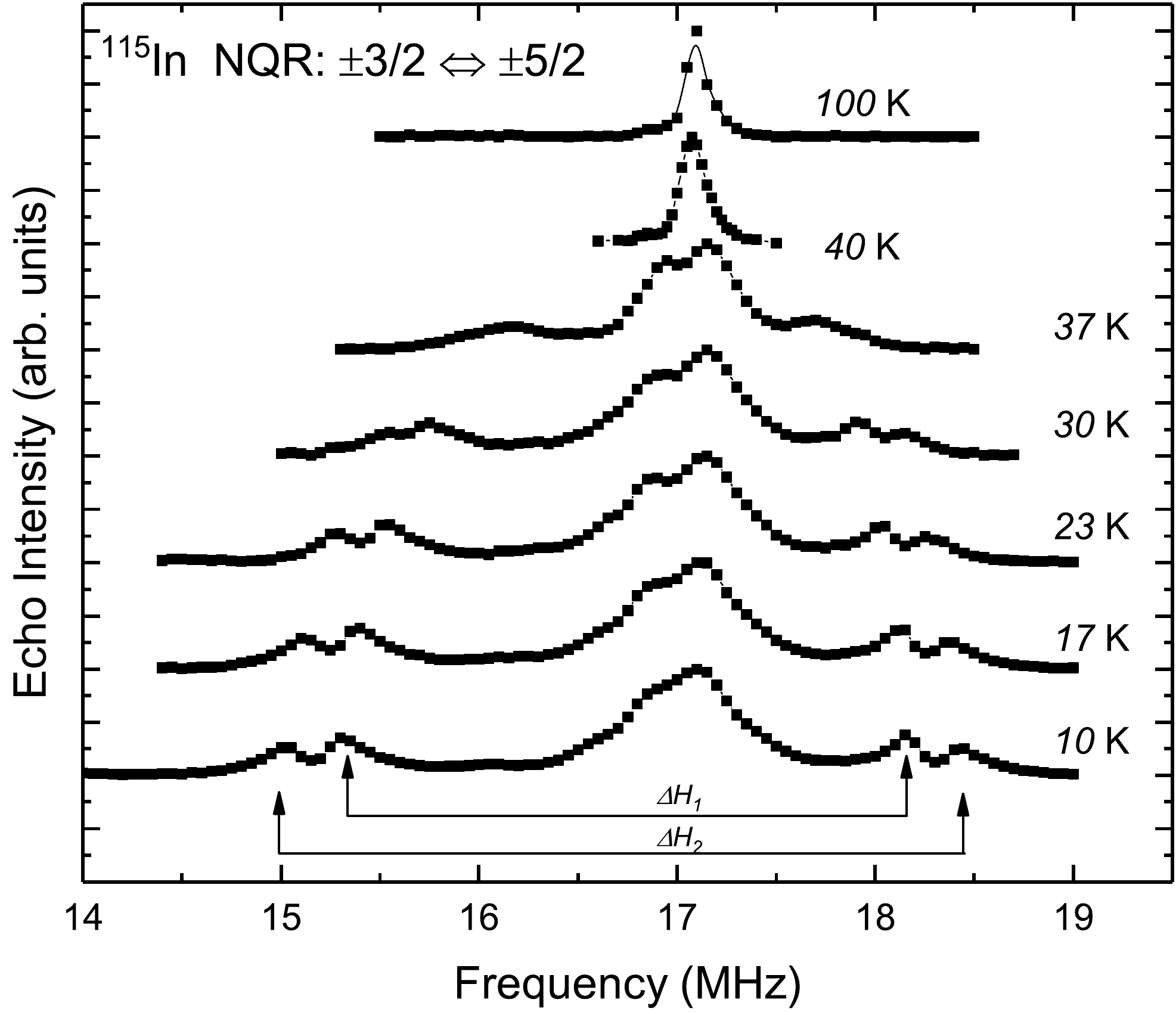}
\caption{ Temperature dependence of the $^{115}$In NQR spectra corresponding to the ($\pm\frac{3}{2} \leftrightarrow \pm\frac{5}{2}$) transition. Arrows labeled $\Delta H_1$ and $\Delta H_2$ indicate pairs of lines split by a strong internal magnetic field. (see the text for details)  } \label{sp}
\end{figure}
%

In the following we focus on the part of the  $^{115}$In NQR spectrum corresponding to the ($\pm\frac{3}{2} \leftrightarrow \pm\frac{5}{2}$) transition which is plotted on an enlarged scale in  Fig.~\ref{HLsp}  for two selected temperatures of 100 and 37\,K. The spectrum measured in the paramagnetic state at 100\,K  consists of a single narrow line at  17.1\,MHz due to an almost perfect overlap of the signals from the In1 and In2 sites for the  ($\pm\frac{3}{2} \leftrightarrow \pm\frac{5}{2}$) transition (see Table~\ref{table1}).  The other spectrum recorded just below the ordering temperature $T^{\ast\ast} = 39$\,K is significantly broader and is split into several lines. As  can be seen in Fig.~\ref{sp} on a larger frequency scale, with further decreasing temperature the splitting of the spectrum  progresses and the lines get more resolved. 

Typically, a splitting of an NQR line of a magnetic material is a fingerprint of an ordered  state. 
If the probed nuclei are located at non-symmetric positions with respect to the AFM sublattices the nuclei are exposed to  local static magnetic fields which cause the Zeeman splitting of the nuclear Kramers doublets (see Appendix for details).
In the symmetrical case the local field from different sublattices  would be compensated and the NQR line should  remain unsplit.  When the spin system is in 
a static but not long range ordered, i.e.,  disordered state, the individual nuclei  in the sample  would sense different local fields  leading to strong line broadening due to  a distribution of local fields \cite{Iakovleva2017}. 

The splitting of the $^{115}$In NQR lines evidences the establishment of the ordered state in \ICVO\ where In nuclei are exposed to local magnetic fields giving rise to magnetically nonequivalent positions at each of the two crystallographic In sites. Apparently, those In nuclei which experience  a stronger local field 
contribute to the lines with a large splitting as indicated in Fig.~\ref{sp} by arrows labeled $\Delta H_1$ and $\Delta H_2$, whereas nuclei experiencing weaker magnetic fields contribute to  the split lines close to the central frequency of 17\,MHz  (see the Appendix).

\subsubsection{$^{115}$In relaxation measurements}
	
\begin{figure}[h]
\includegraphics[width=0.9\columnwidth]{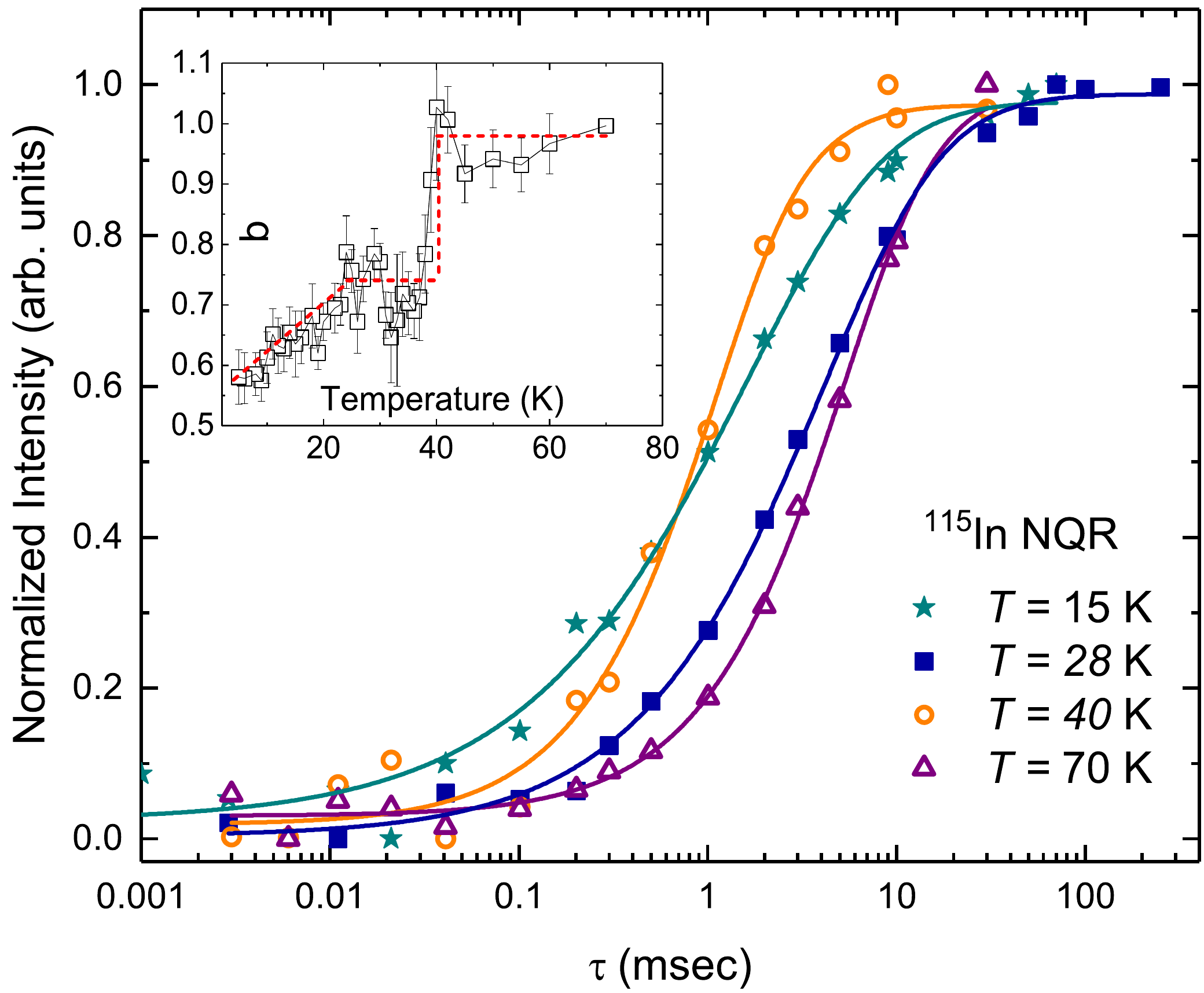}
\caption{(Color online)  Decay of the echo intensity as a function of the time delay $\tau$ between the pulses at temperatures of 70\,K (triangles), 40\,K (circles), 20\,K (squares) and 15\,K(stars). The inset depicts the $T$-dependence of the stretching parameter $b$. (See the text for details).} \label{beta}
\end{figure}

$^{115}$In nuclear spin-lattice relaxation ($T_1^{-1}$) measurements  at the line corresponding to the $\pm3/2\leftrightarrow \pm5/2$ transition  were performed  in a temperature range from 70 down to 5\,K.  In Fig.~\ref{beta}  typical nuclear magnetization recovery curves obtained with the saturation-recovery pulse sequence are presented. 
For nuclei with spin $I>3/2$, the magnetization recovery is a multiexponential function but still  it can be described by a single relaxation time $T_1$. For $I=9/2$, the theoretical magnetization recovery curve contains four exponents. The corresponding  coefficients of the exponential terms were calculated by J. Chepin and J.H. Ross Jr. for the relaxation in magnetic materials as a function of the asymmetry parameter $\eta$ in Ref.~\cite{Chepin}. Since in \ICVO\   two $^{115}$In  sites contribute to the total signal, the following function,  which takes into account the overlap of the two signal contributions, was used to fit the data: 
\begin{align}
M(t)=M_{0}\,[1-(AM_{\rm{In1}}+BM_{\rm{In2}})],\nonumber\\
M_{{\rm{In}}i} = \sum_k C_k{\rm{exp}}({-(\frac{\rho_kt}{T_1})^b}).
\label{t1_fit}
\end{align}
Here $M_0$ is the equilibrium magnetization, $T_1$ is the nuclear spin-lattice relaxation time and $b$ is the stretching parameter. Parameters $A$ and $B$ 
are the weighting factors of the initial nuclear  magnetization after the fist pulse  corresponding to the  In1 and In2 sites, respectively. Since for the transition $\pm3/2\leftrightarrow \pm5/2$ contributions from the two nuclear sites fully overlap these parameters  were taken equal, $A = B = 0.5$. The corresponding numerical coefficients of Eq.~(\ref{t1_fit}) used to fit the magnetization recovery curves for both $^{115}$In positions are given in  Table ~\ref{table2}.   

\begin{table}[h]
\caption{Relaxation exponents $\rho_k$ and corresponding coefficients $C_k$ for the ($\pm\frac{3}{2} \leftrightarrow \pm\frac{5}{2}$) transition taken from Ref.~\cite{Chepin} for $\eta$=0.05 (In1 site) and 0.1 (In2 site).}
\begin{ruledtabular}
\begin{tabular}{c c c c c}

Site  & $\rho_1$  &$\rho_2$& $\rho_3$ &$\rho_4$  \\

In1  & 3 &10  & 20.7 &35.5   \\
In2 & 3 & 10 & 20.5  & 34\\

\colrule

Site &  $C_1$  &$C_2$&$ C_3$ &$C_4$  \\

In1  & 0.016 &0.031  & 0.135 &0.818   \\
In2  & 0.007 & 0.028 & 0.122 &0.843\\
\end{tabular}
\end{ruledtabular}
\label{table2}
\end{table}
\begin{figure}[h]
\includegraphics[width=0.9\columnwidth]{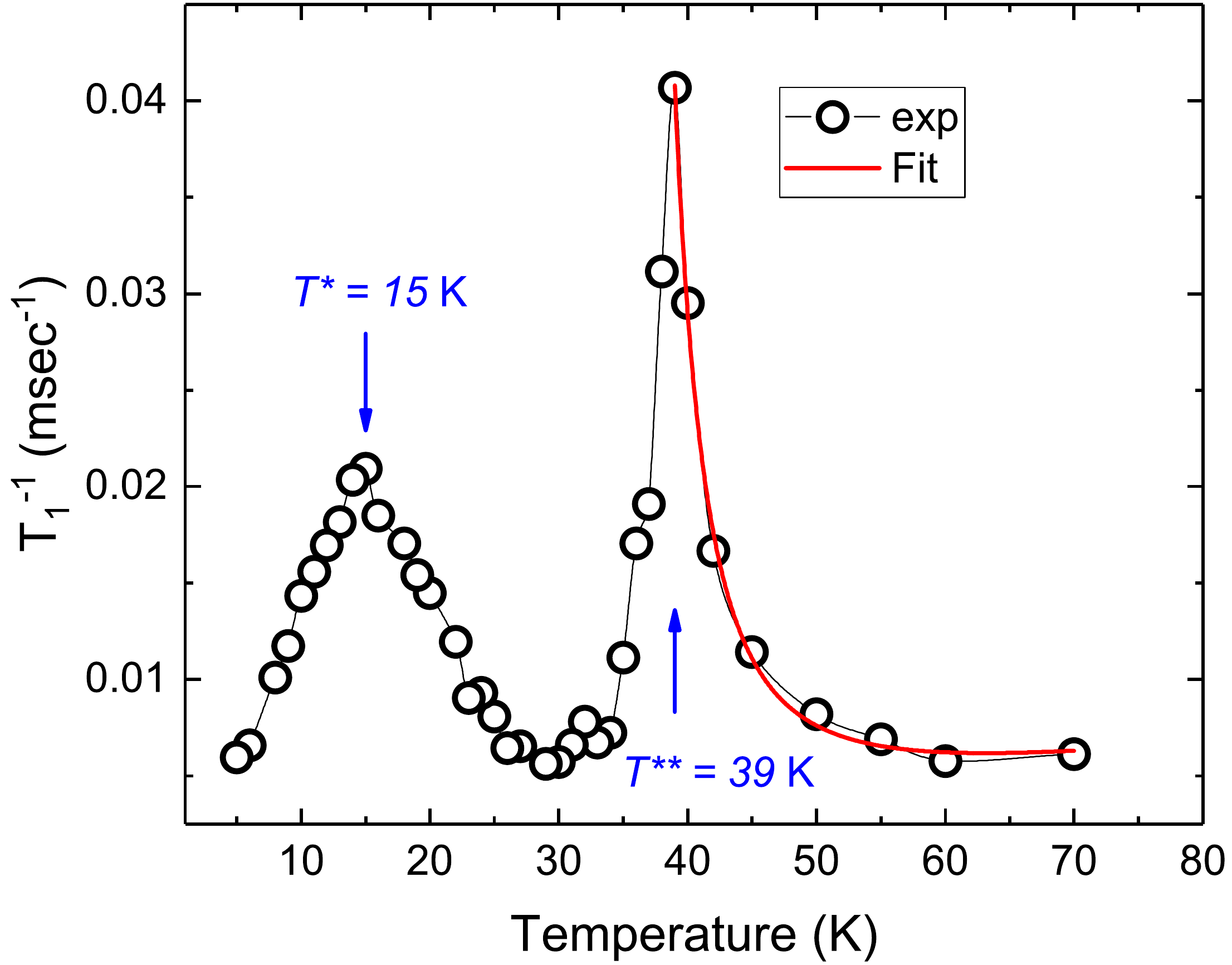}
\caption{Temperature dependence of the $^{115}$In spin-lattice relaxation  rate  $T_1^{-1}$ measured at the $\pm\frac{3}{2}\leftrightarrow\pm\frac{5}{2}$ transition.} 
\label{t1nqr}
\end{figure}

The obtained temperature dependence of the  nuclear spin-lattice relaxation rate $T_1^{-1}$ is shown  in Fig.~\ref{t1nqr}. Generally, $T_1^{-1}$  can be driven  by magnetic fluctuations and/or by a fluctuating EFG  due to lattice vibrations (phonons). The latter mechanism yields $T_1^{-1}$ as a monotonically ascending function of temperature, e.g., $\propto T^2$ or $T^7$ \cite{VANKRANENDONK1954781}, which is not observed experimentally (Fig.~\ref{t1nqr}).
%
%
Furthermore, in Ref.~\cite{PhysRevB.78.024420} three characteristic vibrational modes were observed with temperatures 160, 350 and 710\,K  corresponding to the lattice vibrations at THz frequencies, i.e., far above  the NQR frequency scale.  Thus, such phonon modes should not affect the nuclear relaxation processes as well.  Therefore 
we conclude that the nuclear relaxation is of magnetic origin at low temperatures. Indeed, a gradual increase of the $T_1^{-1}$ rate below 60\,K (Fig.~\ref{t1nqr}) can be ascribed to the development of the magnetic correlations in \ICVO.   Eventually the $T_1^{-1}(T)$ dependence exhibits  a peak at temperature $T^{\star\star}$=39\,K.  Such a sharp peak usually signifies the establishment of  long-range magnetic order. This peak in the relaxation is accompanied by a step-like drop of the stretching parameter $b$,  indicating a distribution of the relaxation times  due to inhomogeneous fluctuating fields in the electron spin system  (Fig.~\ref{beta}, inset). Remarkably, by further decreasing  the temperature the second prominent feature, a broad peak in the $T_1^{-1}(T)$ dependence at  $T^{\star}$=15\,K is observed.  

\subsection{$^{51}$V NMR spin-lattice relaxation rate}

\begin{figure}[b]
	\includegraphics[width=0.9\columnwidth]{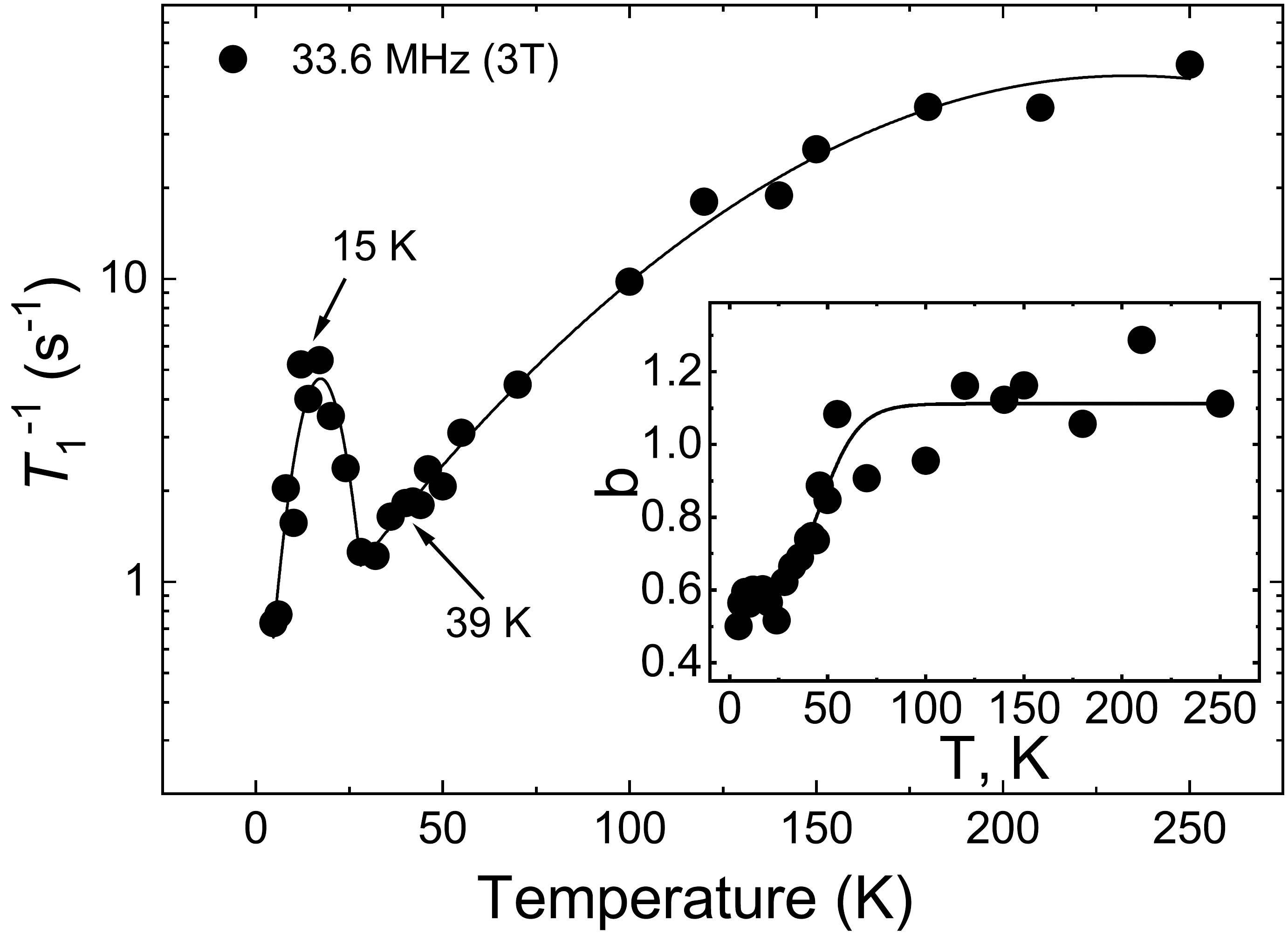}
	\caption{Main panel: Temperature dependence of the $^{51}$V spin-lattice relaxation rate measured in a magnetic field of 3\,T (symbols). Characteristic temperatures $T^\ast = 15$\,K and $T^{\ast\ast} = 39$\,K are indicated by the arrows (cf. Fig.~\ref{t1nqr}).  Inset: $T$-dependence of the stretching parameter $b$ of the decay of the nuclear magnetization (symbols). Solid lines are guides for the eye (see the text for details). } 
	\label{t1nmr} 
\end{figure}
The $^{51}$V NMR spin-lattice relaxation rate $^{51}T_{1}^{-1}$ in  \ICVO\ was measured at the central peak of the $^{51}$V NMR spectrum \cite{PhysRevB.81.060414} in magnetic fields of about 3\,T. The time evolution of the nuclear magnetization $M(t)$ could be described with a single spin-lattice relaxation time $T_1$ in the functional form described by A.~Narath \cite{Narath1967} for the central transition for the nuclear spin $I = 5/2$. With decreasing temperature $^{51}T_{1}^{-1}(T)$ continuously decreases down to $\sim 35$\,K and then exhibits a broad peak centered around $\sim 15$\,K  similar to the peak at $T^\ast =15$\,K in the $T$-dependence of the $^{115}$In NQR relaxation rate (cf. Fig.~\ref{t1nqr}). Notably, in contrast to the $^{115}$In NQR data, no peak at $T^{\ast\ast} =39$\,K 
was found in the $T$-dependence of the $^{51}$V NMR relaxation rate. The inset of Fig.~\ref{t1nmr} depicts the dependence of the stretching parameter $b$ of the nuclear magnetization decay on temperature. $b$ is close to unity at high temperatures and, similar to the case of the $^{115}$In NQR relaxation (cf. Fig.~\ref{beta}, inset), rapidly decreases down to $b\sim 0.5$ below 40\,K.

\subsection{$\mu$SR experiments}

\subsubsection{Transverse field $\mu$SR measurements}

\begin{figure}[h]
	\includegraphics[width=1.1\columnwidth]{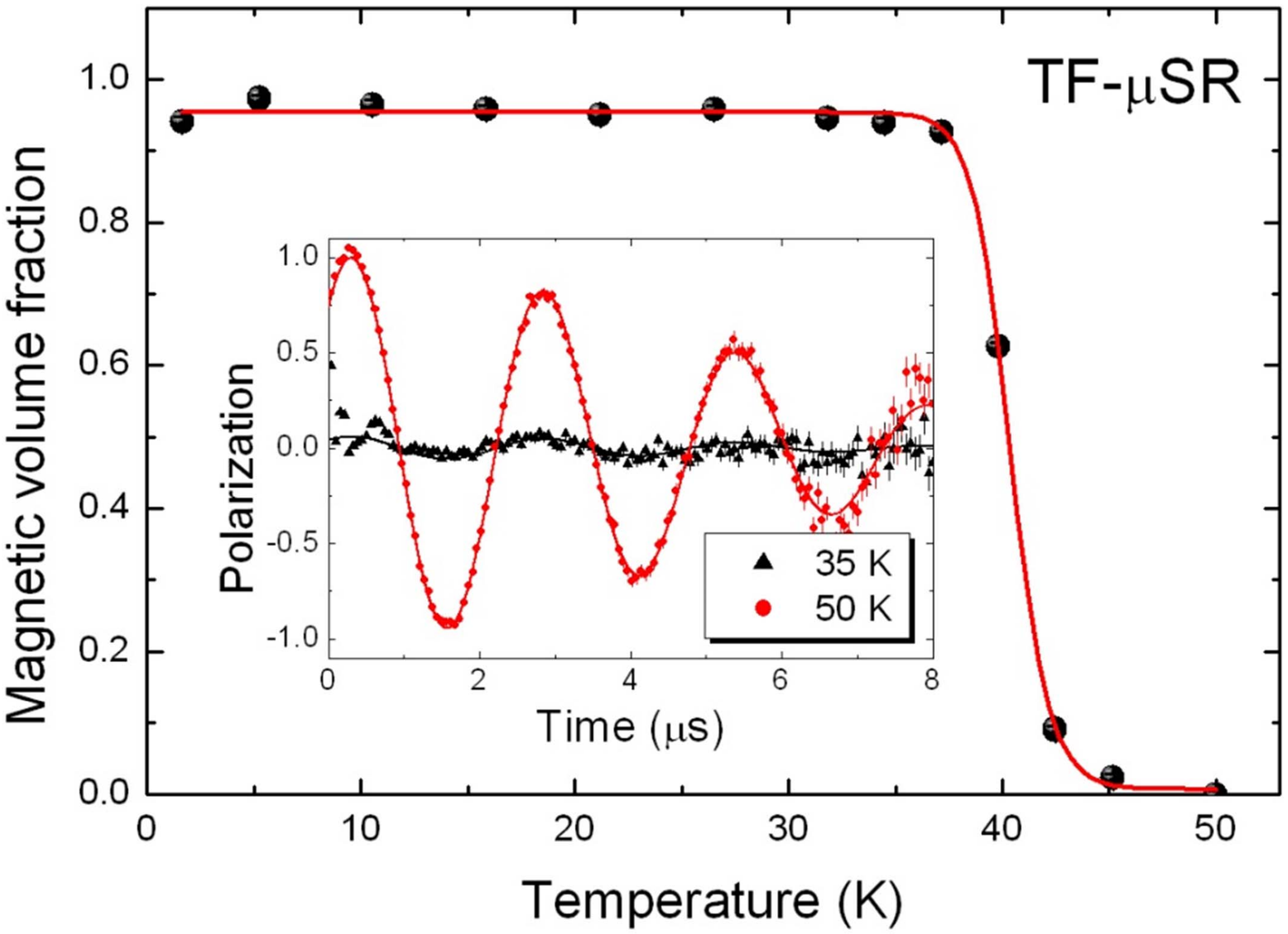}
	\caption{Transverse field $\mu$SR results of \ICVO. Main panel: Temperature dependence of the magnetic volume fraction determined from the amplitude fraction of muons spins precessing with the Larmor frequency $\omega_{\mu} = \gamma_{\mu}\, B_{\mathrm {ext}}$ corresponding to the external field. The solid line is a guide for the eye. Inset: Typical time-dependent muon spin polarization at 60 K (red dots) and at 35~K (black triangles). The solid lines are results of a least-square non-linear fit analysis  of the data as described in the text.}
	\label{TFmusr} 
\end{figure}

The results of the transverse field (TF) $\mu$SR measurements in 5\,mT external field are depicted in Fig.\,\ref{TFmusr}.
At 50\,K the full muon spin polarization amplitude is oscillating with a frequency corresponding to the external field. A weak depolarization is due to a weak static field distribution at the muon site of the order of 10\,G caused by nuclear dipole moments. At 35\,K the amplitude of this signal is reduced to 5\,\% and 95 \,\% of the signal show a much faster depolarization due to the appearance of strong internal magnetic fields at the muon site which exceed the external field by more than a factor of 10. The zero field $\mu$SR measurements discussed below prove that these internal fields  are static on the timescale of microseconds.  
The muon spin polarization $P(t)$ in the TF time spectra was analyzed using the polarization function
\begin {equation} 
P(t)= P_0 (1 - f_{\mathrm {mag}}(T)) \cos{(\omega_{\mu} t + \phi)} exp {(-\sigma_{\mathrm {TF}} t)^2}.
\end{equation}
Here, $f_{\mathrm {mag}}(T)$ is the fraction of muons experiencing a strong internal magnetic field due to magnetic order, 
$\phi$ is the initial phase of the muon spin
polarization at $t = 0$ and  $\sigma_{\mathrm {TF}}$ is the static Gaussian line width at 60\,K. 
The main panel of Fig.~\ref{TFmusr} shows $ f_{\mathrm {mag}}(T)$. For temperatures below 40\,K, $ f_{\mathrm {mag}}(T)$ exhibits a nearly constant value of 0.95. The remaining precession signal amplitude can be associated with muons stopped in the sample holder. Therefore, the TF $\mu$SR measurements prove that muons in the full sample volume experience strong static internal fields below 40\,K.

\subsubsection{Zero field $\mu$SR measurements}

\begin{figure}[h]
	\includegraphics[width=1.1\columnwidth]{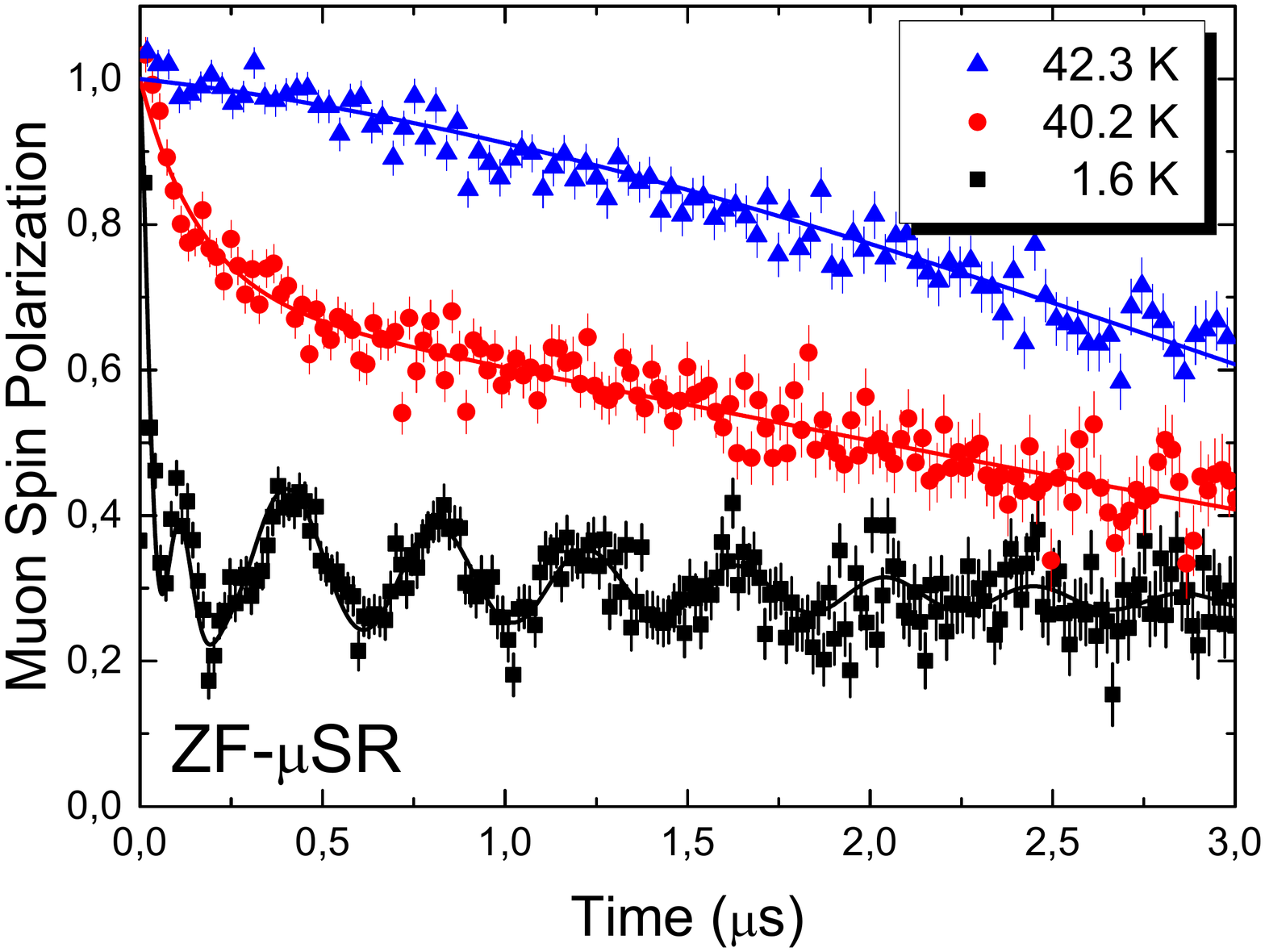}
	\caption{ Typical zero field $\mu$SR time spectra of \ICVO{} at 42.3 K (blue triangles), 40.2 K (red dots) and 1.6~K (black squares). The solid lines are results of a least-square non-linear fit analysis  of the data as described in the text.} 
	\label{ZFspectra}
	\end{figure}

Zero field (ZF) $\mu$SR measurements were performed between 2\,K and 42.5\,K. Typical time spectra are depicted in Fig.~\ref{ZFspectra}.
At 42.5\,K the muon spin polarization decays only weakly with a Gaussian time dependence. The relaxation rate $\sigma_{\mathrm {ZF}}$ is consistent with the TF measurements at high temperatures.
At 40.2\,K the onset of strong relaxation of a fraction of the full signal amplitude sets in.
At 1.6\,K two different spontaneous precession frequencies can be resolved, a weakly relaxing component with frequency $\nu_{1} \approx$ 2.5 MHz and strongly relaxing component with  $\nu_{2} \approx$ 8 MHz. 
A fraction of $f_{\mathrm {long}} \approx 1/3$ of the initial muon spin polarization  does show only a weak relaxation in the microsecond time range without oscillations.  This proves that (i) two different muon sites are populated in  \ICVO\ most likely due to OH-like bonds of the muon to different oxygen ions in the lattice which are energetically nearly degenerated,  and (ii) that the internal fields at both muon sites are nearly static. Therefore the ZF muon spin polarization $P(t)$
below 40\,K was analyzed using the polarization function
\begin{flalign*} 
&P(t) =  P_0[f_\mathrm {long} \, \exp {(- \lambda_\mathrm {l} \, t)} + 
        (1 -  f_\mathrm {long}) \, \exp{(- \sigma_{\mathrm {ZF}} \, t)^2}& \\ 
&       [ f_\mathrm {l}(T) \cos{(\omega_{\mathrm {l}}(T)\, t}) \, \exp {(- \lambda_{\mathrm {t},1}(T)\, t)}\, + & \\
&      (1- f_\mathrm {l}(T)) \cos{(\omega_{\mathrm {2}}(T)\, t }) \, \exp {(- \lambda_{\mathrm {t},2}(T)\, t)}]].
\end{flalign*}
Here, $f_{1}(T)$ is the relative fraction of muon site 1 exhibiting the lower muon spin precession frequency $\nu_{1} = \omega_{1}/2 \pi \approx$ 2.5 MHz with the exponential relaxation rate $\lambda_{\mathrm {t},1}$, and $\nu_{2} =\omega_{2}/2 \pi \approx$ 8 MHz with the exponential relaxation rate $\lambda_{\mathrm {t},2}$ being the corresponding parameters of muon site 2. The relative
signal amplitude of muon site 1,  $f_{1}(T) \approx$\,0.25 is temperature independent below 30\,K. Above 30~K it increases towards 0.6 close to 40\,K. Experimentally it was difficult to follow the high frequency ($\nu_2$) signal close to $T_{\rm N}$ due to its high relaxation rate. Therefore close to $T_{\rm N}$ the fitting algorithm increases the signal fraction of the low frequency ($\nu_1$)  signal (which, of course, also exhibits a frequency approaching zero close to $T_{\rm N}$ and allows some ambiguity in amplitude). Since we cannot exclude a change of the relative muon site occupancies ( due to small  changes of the binding potential energies at the two muon sites) we did not fix this parameter in the analysis.

\begin{figure}[h]
	\includegraphics[width=1.1\columnwidth]{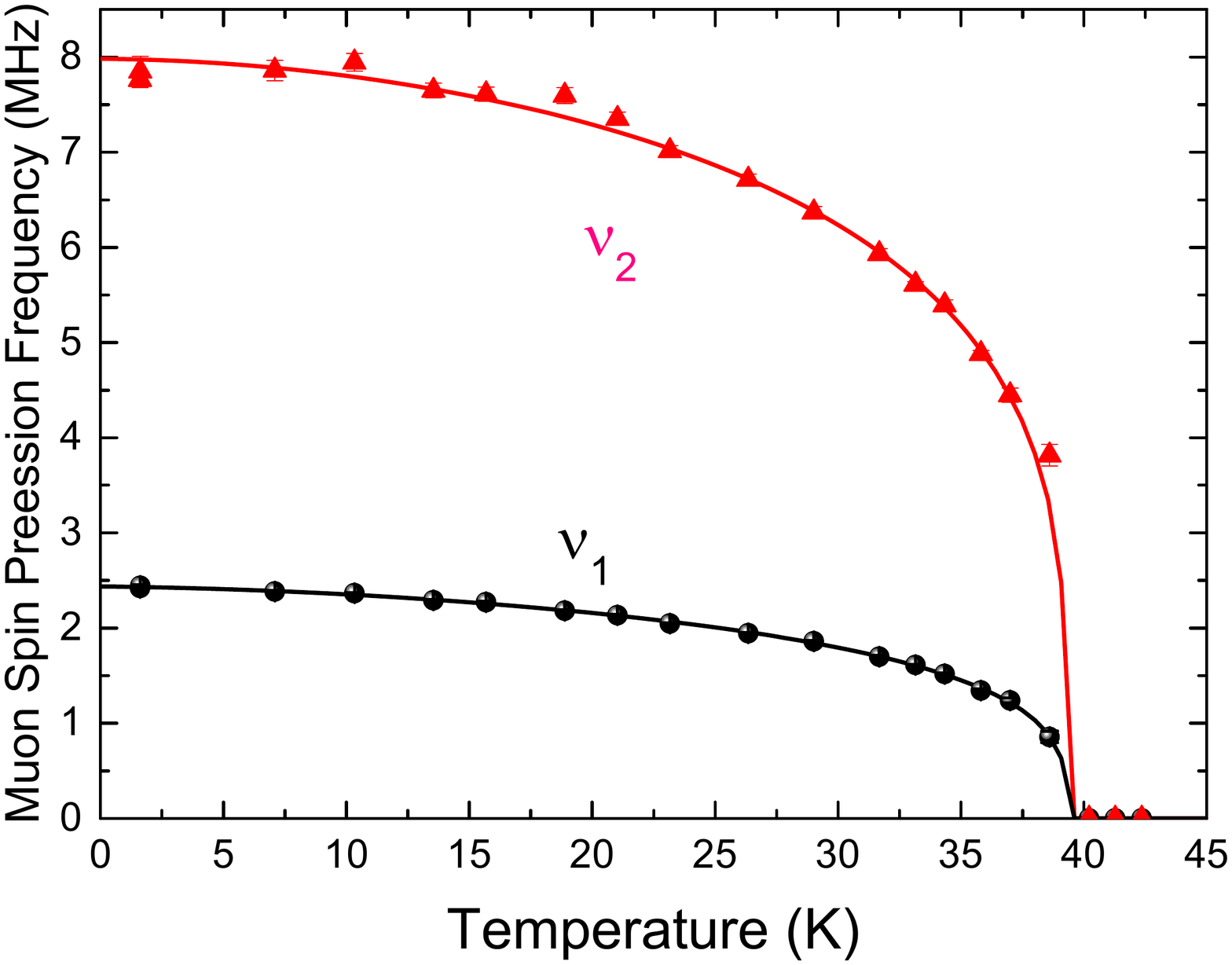}
	\caption{Temperature dependence of the spontaneous muon spin precession frequencies $\nu_1$ (black dots) and $\nu_2$ (red triangles) in the magnetically ordered phase of \ICVO{} below 40\,K. The global fit using an order parameter function Eq.~(\ref{orderparam}) with $T_\mathrm {N}$\,=\,39.1(1)\,K and $\beta $= 0.25(1)  is described in the text.  } 
	\label{ZFfrequency}
	\end{figure}

In Fig.~\ref{ZFfrequency} the temperature dependence
of both muon spin precession frequencies is shown.
%
%
In a global fit  $\nu_{1}$  and $\nu_{2}$ are very well described by an order parameter function of the form
\begin {equation} 
\nu_{i}(T) = \nu_{i,0} \, (1- T/T_N)^{\beta},\, (i=1,2).
\label{orderparam}
\end{equation}
Here, $T_\mathrm {N}$\,=\,39.1(1)\,K and the critical exponent $\beta $= 0.25(1) are global parameters, and  $\nu_{1}$\,= 2.45(1)\,MHz
and $\nu_{2}$\,= 7.98(1)\,MHz. Clearly both curves show no anomaly near $T^\ast = 15$\,K, whereas $T_{\rm N}$ is equal to $T^{\ast\ast} = 39$\,K (cf. Fig.~\ref{t1nqr}).

\begin{figure}[h]
	\includegraphics[width=0.91\columnwidth]{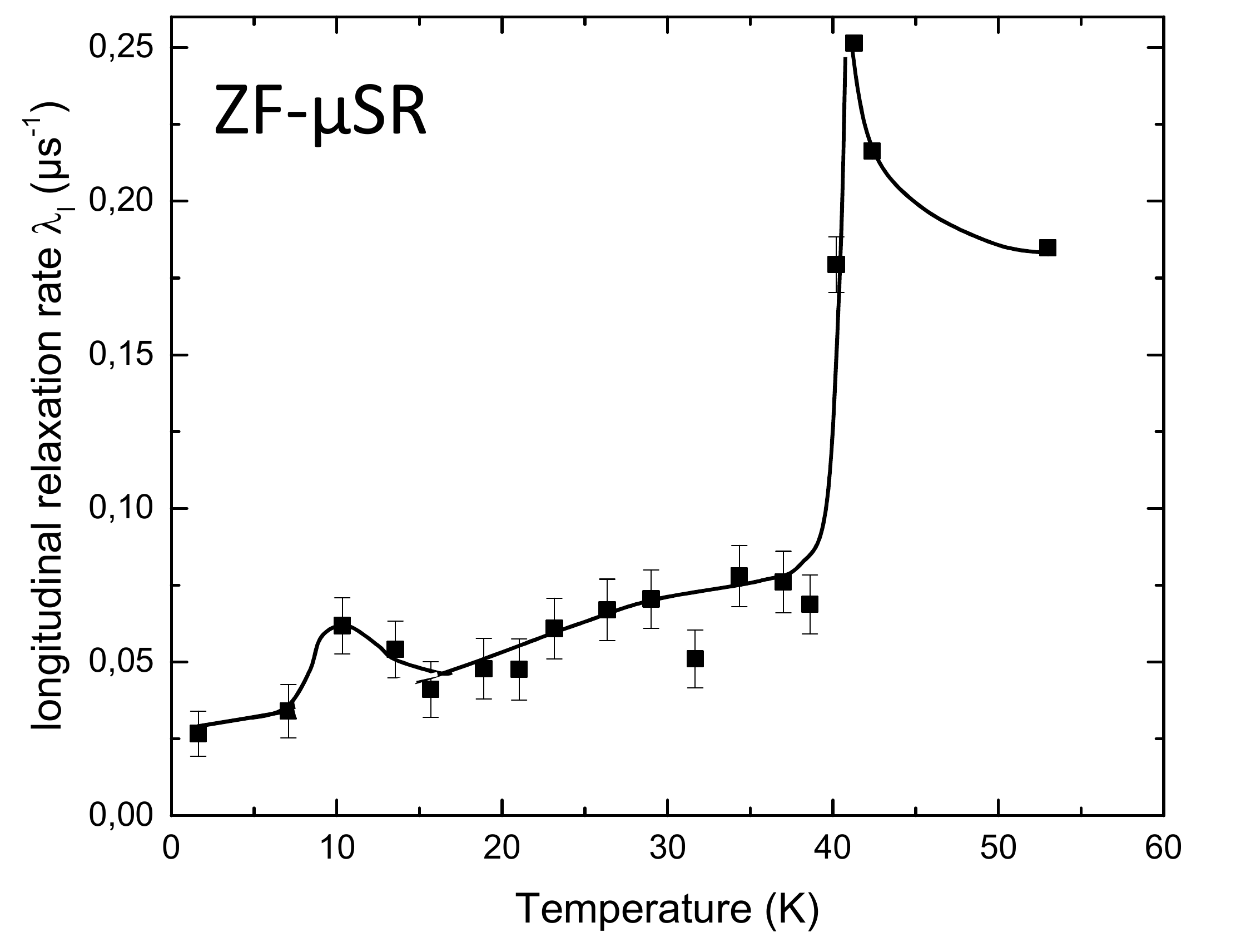}
	\caption{Temperature dependence of the longitudinal relaxation rate $\lambda_{\mathrm {l}}$ in \ICVO. In the magnetically ordered phase below 40\,K in general a weak monotonous decrease of $\lambda_{\mathrm {l}}$ is observed superimposed by a small peak at $\approx$\,10\,K. The solid and dashed lines are guides to the eye.} 
	\label{ZFlamlong}
	\end{figure}

 The longitudinal relaxation rate  
  $\lambda_{\rm {l}}$,  depicted in Fig.\, \ref{ZFlamlong}, exhibits a maximum close to $T_{\rm N}$.
Below 38\,K a slow decrease is observed from 0.07 $\mu s^{-1}$  towards 0.03 $\mu s^{-1}$ at 2\,K. Only a weak second maximum is found between 10\,K and 15\,K reminiscent
of the broad peak observed in $T_1^{-1}$ of $^{115}$In NQR at $T^\ast = 15$\,K. The temperature shift of the peak position in the $\mu$SR experiment is consistent with the fact that the spontaneous muon spin precession frequencies are smaller than the NQR transition frequency by a factor of $\approx$ 2 -- 6 depending on the muon site.    
	
\begin{figure}[h]
	\includegraphics[width=1.1\columnwidth]{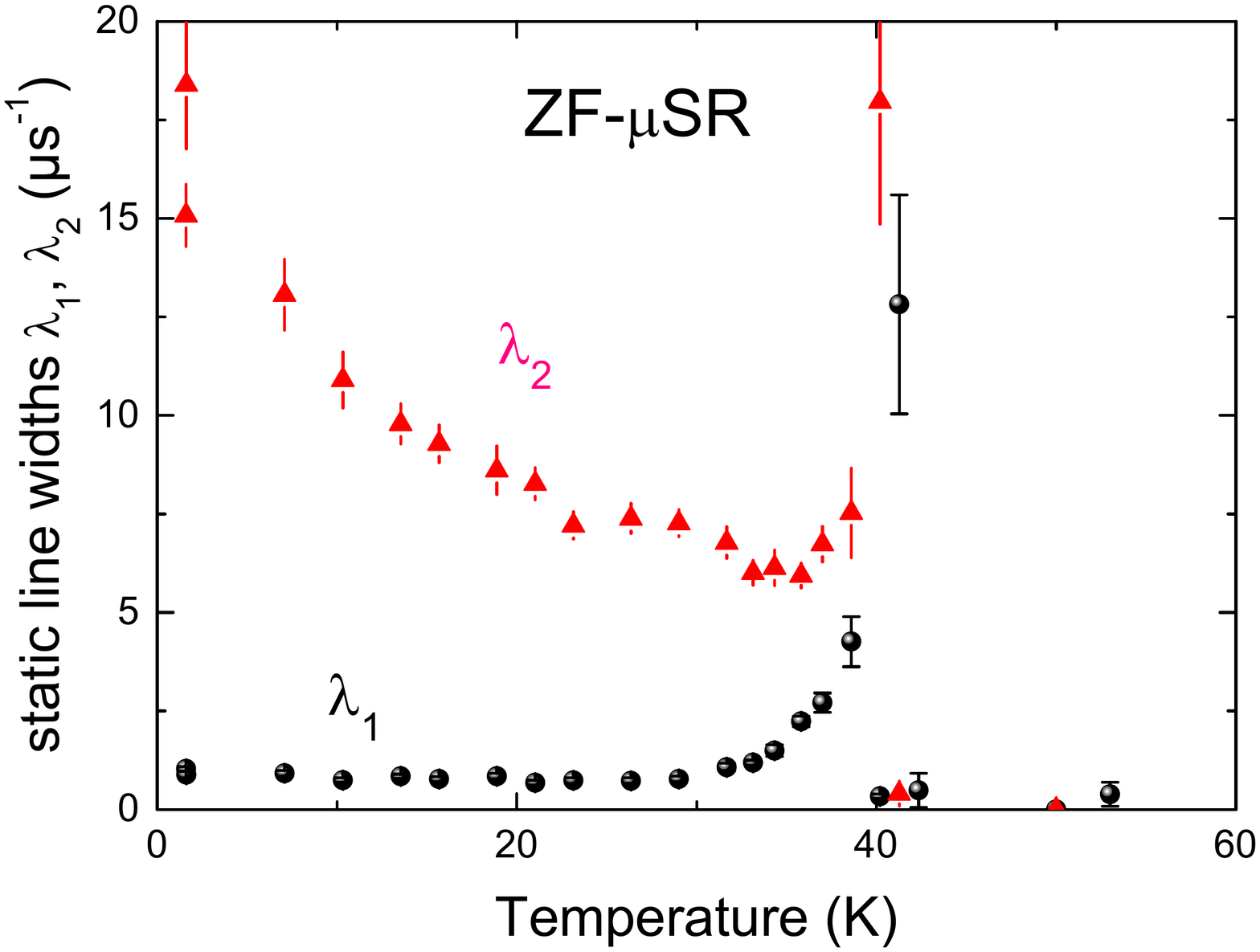}
	\caption{ Temperature dependence of the static line widths $\lambda_{\mathrm {t},1}$ (black dots) and $\lambda_{\mathrm {t},2}$ (red triangles) in the magnetically ordered phase of \ICVO{} below 40\,K.}
	\label{ZFwidths}
	\end{figure}

The essentially static line widths  $\lambda_{\rm {t},1}$ and $\lambda_{\rm {t},2}$ are depicted in Fig.\, \ref{ZFwidths} (note that $\lambda_{\rm {l}}$ is $1-2$ orders of magnitude smaller). Since the local magnetic field at site 2 is $\approx$ 4\,times larger than at site 1, $\lambda_{\rm {t},2}$ is also larger than $\lambda_{\rm {t},1}$ by a similar factor. Above 35\,K towards $T_{\rm N}$ both line widths increase due to the reduction of the in-plane magnetic coherence length close to $T_{\rm N}$. Below 30\,K $\lambda_{\rm {t},1}$ is nearly temperature independent whereas $\lambda_{\rm {t},2}$ increases towards lower temperatures. This increase in linewidth may result from the onset of the static interlayer correlations in this temperature range. 

\section{Microscopic model}\label{model}
Scalar relativistic DFT calculations have been performed within the generalized
gradient approximation (GGA)~\cite{oj:PBE96} as implemented in the full
potential code \textsc{fplo} version 18~\cite{oj:FPLO}. Nonmagnetic band
structure calculations were performed on a 14$\times$14$\times$7 $k$ mesh (228
points in the irreducible wedge). For the structural input, we used the neutron
powder diffraction data measured at 10\,K~\cite{PhysRevB.78.024420}.  Note that
the lattice constants and atomic coordinates provided in Table~I of
Ref.~\cite{PhysRevB.78.024420} pertain to the Cu/V disordered structure with
the space group $P6_3/mmc$ (194) and can not describe the honeycomb lattice
structure of \ICVO. The highest symmetry compatible with the honeycomb
arrangement of Cu atoms for a single layer is $P\bar{6}2m$ (189).  Yet it features a simple
stacking of magnetic honeycomb planes without a shift, giving rise to the
spurious trigonal prismatic local coordination of In. Thus, the minimal
structural model of \ICVO\ entails the A-B-A-B stacking, where the A and B
planes are shifted with respect to each other, leading to the space group
$Cmcm$ (63).

The lack of a threefold rotation symmetry in the orthorhombic $Cmcm$ structure
singles out one of the three nearest-neighbor exchanges, which is not anymore
equivalent to the remaining two, and raises the question of a possible
dimerization. To estimate this tendency, we relaxed the oxygen positions within
the GGA+$U$ with $U_d$\,=\,8.5\,eV and $J_d$\,=\,1\,eV using the fully
localized limit as the double counting correction. To allow for an
antiferromagnetic arrangement within the honeycomb planes, the symmetry has
been further lowered down to monoclinic, space group $P2_1/m$ (11). The
resulting structure predictably has a lower GGA energy, but the difference
between the transfer integrals for the inequivalent paths is small: $-174$\,meV
for the ``singled out'' path versus $-180$\,meV for the other two paths.
Moreover, since the former transfer integral is smaller, and the respective
antiferromagnetic exchange is weaker, indicating that there is no tendency
towards an electronically-driven dimerization in the honeycomb planes.
This also agrees with the experimental data where no fingerprints of a
dimerization have been observed. Therefore, all further calculations were
performed for the experimental crystal structure.

\begin{figure}[tb]
\includegraphics[width=\columnwidth]{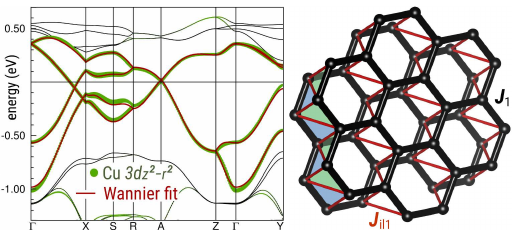}
\caption{\label{fig:dft} Left panel: GGA band structure of \ICVO. The radii
of the green circles denote the weight of the half-filled Cu $d_{3z^2-r^2}$
orbital. Red solid lines are eigenvalues of the tight-binding Hamiltonian
constructed from the Wannier projections. The notation of the $k$ points is:
$\Gamma$\,=\,$[0,0,0]$, X\,=\,$[0,\frac13,0]$, S\,=\,$[-\frac14,\frac14,0]$,
R\,=\,$[-\frac14,\frac14,\frac12]$, A\,=\,$[0,\frac13,\frac12]$,
Z\,=\,$[0,0,\frac12]$, and Y\,=\,$[-\frac12,0,0]$ in terms of the reciprocal
lattice vectors of the conventional ($C$-centered) unit cell. Right panel: a
sketch of the spin model of \ICVO\ with the nearest-neighbor exchange $J_1$
(black cylinders) and the interlayer exchange $J_{\text{il1}}$ (red
lines). Both exchanges are antiferromagnetic.  Shaded triangles illustrate
the magnetic frustration and are a guide to the eye. The spin model picture has
been created using \textsc{vesta}~\cite{oj:vesta}.
}

\end{figure}

Though in the chemical notation \ICVO\ is named indium copper oxide-vanadate, from the physics perspective, electronically it is an undoped cuprate with the $3d^9$ electronic
configuration.
Typical for this class of materials, GGA calculations yield a
metallic ground state due to the severe underestimation of electronic
correlations. The four Cu atoms in the unit cell produce a four-band manifold
crossing the Fermi level (Fig.~\ref{fig:dft}, left). In contrast to most
cuprates that feature the half-filled Cu $3d_{x^2-y^2}$ orbital, this band
manifold in \ICVO\ corresponds to the antibonding combination of
$\sigma$-overlapping Cu $3d_{3z^2-r^2}$ and O $2p_{z}$ orbitals. The strong
hybridization allows us to resort to an effective single-orbital model with one
orbital per Cu. The transfer integrals $t_{ij}$ between these effective
orbitals are estimated by Wannier projections~\cite{oj:eschrig09}. In this way,
we find that only three terms exceed 10\,meV: the first and second neighbors in
the honeycomb plane --- $t_1$, and $t_2$, respectively, --- as well as the
shortest interlayer coupling $t_{\text{il}1}$.

The antiferromagnetic exchange can be directly estimated in second order
perturbation theory as $J^{\text{AF}}_{ij}$\,=\,$4t_{ij}^2/U_{\text{eff}}$,
where $U_{\text{eff}}$ is the onsite Coulomb repulsion within the effective
one-orbital model. By taking the commonly used value
$U_{\text{eff}}$\,=\,4.5\,eV (e.g.~\cite{oj:janson09}), we obtain the leading
antiferromagnetic exchange $J^{\text{AF}}_1$\,=\,360\,K, which is amenable to a
direct comparison with the experiment:  The magnetic susceptibility of the
$S$\,=\,$\frac12$ honeycomb Heisenberg model has a broad maximum centered at
$\sim$0.72$J_1$~(e.g., \cite{oj:tsirlin10}).  By taking the experimental position
of this maximum (185\,K~\cite{PhysRevB.78.024420}) in \ICVO, we obtain $J_1$ of
about 255\,K.  The reduced value of $J_1$ hints at a sizable ferromagnetic
contribution to the magnetic exchange, which is lacking in the effective
one-orbital approach in accord with the Pauli principle.

Next, we estimated the total exchange integrals, containing both
antiferromagnetic and ferromagnetic contributions.  To this end, we performed
spin-polarized calculations, using a supercell doubled along the $a$ axis and
calculated the total energies within the GGA+$U$ approach with the Coulomb
repulsion $U_d$\,=\,8.5\,eV, the Hund's exchange $J_d$\,=\,1.0\,eV, and the
fully localized limit~\cite{oj:czyzyk95} as the double counting correction.
All calculations have been done in the cell metrically equivalent to the unit
cell doubled along the $a$ axis, the space group $Pm$ (6), on a mesh of
$2\times4\times4$ $k$ points.  As expected, the GGA+$U$ restores the insulating
nature of \ICVO, while the orbital occupation matrices indicate that the
half-filled (and hence, magnetically active) orbital in \ICVO\ is Cu
$3d_{3z^2-r^2}$.

The GGA+$U$ total energies of 31 different collinear magnetic configurations
were mapped onto a classical Heisenberg model with $|\vec{S}_i|$\,=\,$\frac12$.
The five short-range (with $d_{\text{Cu..Cu}}$ up to 7.5\,\r{A}) exchange
integrals $J_1$, $J_2$, $J_3$, $J_{\text{il}1}$, and $J_{\text{il}2}$ were
determined by a least-squares solution to a redundant linear problem; the
results are provided in the last column of Table~\ref{tab:exchanges}. 

A corollary of this analysis is the presence of the dominant nearest-neighbor
exchange $J_1$ and the much weaker interlayer exchange $J_{\text{il}1}$ (see
the right panel of Fig.~\ref{fig:dft} for a sketch of the model), while further
exchange couplings are comparable to the error bars. At this point, it is
crucial to consider the topology of the spin lattice. As we discussed earlier
in this Section, the neighboring layers in \ICVO\ are shifted with respect to
each other. Due to this shift, two interplane exchanges couple the nearest
neighbors of one plane with the same spin in the neighboring plane. The
resulting $J_1$--$J_{\text{il}1}$--$J_{\text{il}1}$ triangles underlie the
geometrical frustration of the magnetic model~(Fig.~\ref{fig:dft}, right).  We
conclude that \ICVO\ is an excellent realization of the honeycomb lattice where
the magnetic ordering is further suppressed by a frustrated interlayer
exchange.

Let us note that the absolute numerical values of $J_1$ and $J_{\text{il}1}$
may be up to $15$\,\% inaccurate due to the ambiguous choice of the $U_d$
parameter. Yet their ratio $J_{\text{il}1}/J_1\simeq2.5$\,\% and the
irrelevance of further exchanges beyond the $J_1$--$J_{\text{il}1}$ model are a
solid outcome of the analysis. It is clear that in the J$_{\text{il}1}\ll{}J_1$
regime, the anisotropy of $J_1$ may play a crucial role as its magnitude can be
larger than the isotropic exchange $J_{\text{il}1}$. To investigate the
anisotropic terms, we perform full-relativistic noncollinear GGA+$U$
calculations using \textsc{vasp} version 5.4.4~\cite{oj:VASP, *oj:VASP_2} with
projector-augmented-wave pseudopotentials~\cite{oj:bloechl94,oj:VASP_pseudo}.
Anisotropic exchange parameters were evaluated by using the four-cell
method~\cite{oj:xiang13}. The resulting bilinear exchange tensor $M_1$ is:

\begin{equation}
  \label{eq:j1}
  M_1 = \left(\begin{array}{ddd}239.4 & 0 & 0.7 \\ 0 & 230.1 & 0 \\ -0.7 & 0 &
240.4\end{array}\right) \text{K},
\end{equation}

where the honeycomb planes are in the $xz$ plane. Thus, the antisymmetric part
of $M_1$ describes the Dzyaloshinksii-Moriya vector $|D_1|$\,=\,0.7\,K
perpendicular to the honeycomb planes. The extremely small value of $|D_1|/J_1$
of $\sim$0.3\,\% again indicates that \ICVO\ is an
excellent model honeycomb system. In contrast, the diagonal elements reveal a
considerable XXZ anisotropy of $\sim$4\,\%.

\begin{table}[tb]
  \caption{\label{tab:exchanges} Transfer integrals $t_{ij}$ (in meV) of the
effective one-orbital model and the respective magnetic exchanges $J_{ij}$ (in
K) evaluated based on total energy GGA+$U$ calculations.  For each magnetic
exchange, the respective interatomic Cu..Cu distance (in \r{A}) and the
multiplicity (within in the unit cell) are provided.}
  \begin{ruledtabular}
    \begin{tabular}{lccrd}
      path & $d_{\text{Cu..Cu}}$ & multiplicity & $t_{ij}$ & \ensuremath{J_{ij}} \\ \hline
      $X_1$            & 3.3509 & 12 & $-187$ & 211.7 \\
      $X_2$            & 5.8038 & 24 &  $-20$ &  -0.8 \\
      $X_{\text{il}1}$ & 6.2572 & 16 &  $-49$ &   5.4 \\
      $X_3$            & 6.7018 & 12 &     6  &   0.3 \\
      $X_{\text{il}2}$ & 7.0979 & 16 &     2  &  -0.6 \\
    \end{tabular}
  \end{ruledtabular}
\end{table}

\begin{figure}[!t]
	\includegraphics[width=8.6cm]{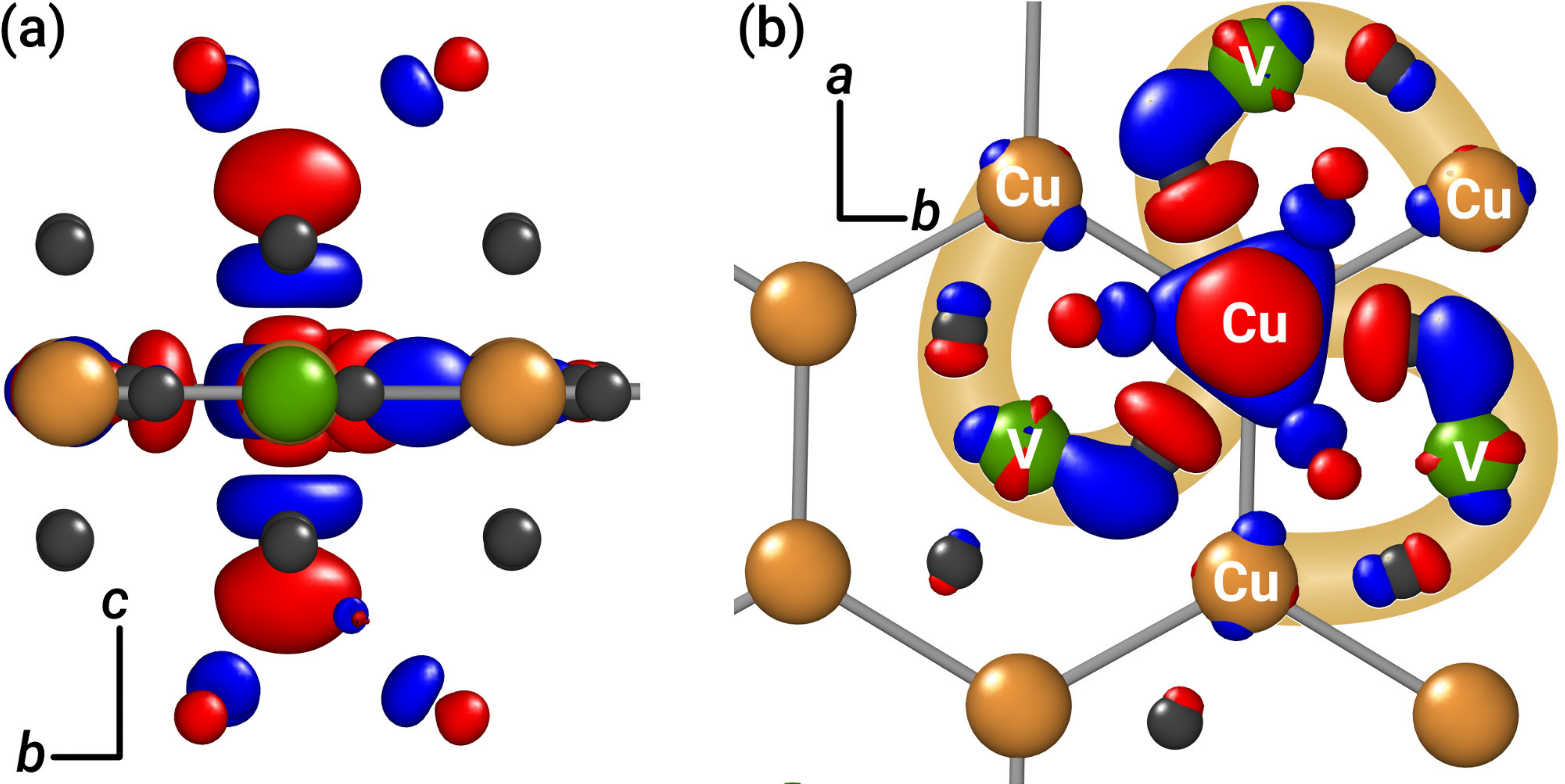}
	\caption{\label{fig:wf} Cu-centered Wannier functions for the $|3z^2-r^2\rangle$ states in \ICVO: (a) lateral and (b) top views. The Cu--O--V--O--Cu paths facilitating a sizable antiferromagnetic $J_1$ exchange are highlighted in the right plot.
	}
\end{figure}

The crystal field generated by the trigonal bipyramidal environment of copper
atoms in \ICVO\ renders the $|3z^2-r^2\rangle$ orbital half-filled and
magnetically active. This unusual orbital ground state has been previously
conjectured for two other cuprate materials: the spin chain system
CuSb$_2$O$_6$~\cite{oj:heinrich03, oj:kasinathan08} and the 3D skyrmionic Mott
insulator Cu$_2$OSeO$_3$~\cite{oj:seki12, oj:janson14}. 

In CuSb$_2$O$_6$, the CuO$_6$ octahedra are squeezed, forming two short and four
long Cu--O bonds. DFT calculations indicate a small crystal field splitting,
but the $|3z^2-r^2\rangle$ orbital becomes half-filled due to the larger band
width, which in turn gives rise to a sizable gain in kinetic energy~\cite{oj:kasinathan08}.  Hence, the orbital ground
state is stabilized by the competition between hopping processes and the onsite
Coulomb repulsion rather than by the crystal field.

The case of Cu$_2$OSeO$_3$ is closer to \ICVO: here, one of the two
structurally inequivalent copper atoms, Cu(1), has the local trigonal
bipyramid environment.  The microscopic magnetic model features five
inequivalent exchanges, two ferromagnetic exchanges connecting Cu(2)
atoms that have the conventional $|x^2-y^2\rangle$ orbital ground state, and
three antiferromagnetic exchanges that couple Cu(1) and Cu(2) sublattices.
Interestingly, one of these antiferromagnetic exchanges is accompanied by a
large Dzyaloshinskii-Moriya anisotropy amounting to 58\% of the
isotropic exchange~\cite{oj:janson14}. However, in \ICVO\ the situation is
remarkably different: the $|3z^2-r^2\rangle$ orbitals of the neighboring atoms
are stretched perpendicular to the respective nearest-neighbor bonds, and the
Dzyaloshinskii-Moriya anisotropy is nearly absent. 

Coming back to the isotropic model, \ICVO\ has a sizable nearest-neighbor
superexchange, while longer-range intraplane exchanges $J_2$ and $J_3$ are
strongly suppressed.  As can be seen from the Wannier functions of
$|3z^2-r^2\rangle$ states in Fig.~\ref{fig:wf}, this behavior can be understood
as a joint effect of the $|3z^2-r^2\rangle$ orbital state and the strong
covalency of V-O bonds.  A sizable  V-mediated superexchange is present in
other low-dimensional magnets, such as volborthite
Cu$_3$V$_2$O$_7$(OH)$_2{\cdot}2$H$_2$O~\cite{oj:janson16} and
SrNi$_2$V$_2$O$_8$~\cite{oj:bera15}.

\section{Discussion}\label{Discussion}

\begin{figure}[h]
	\includegraphics[width=1.0\columnwidth]{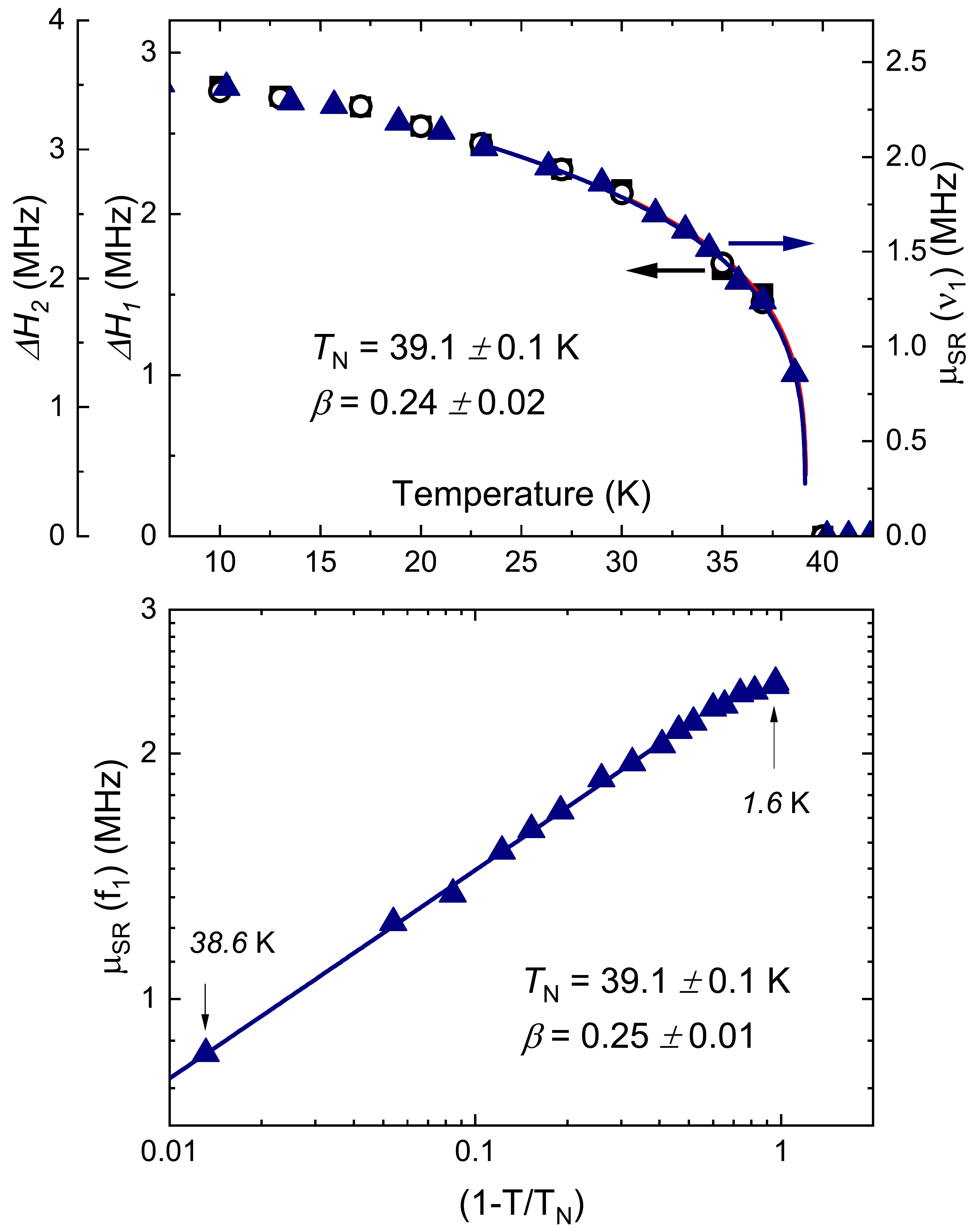}
	\caption{Upper panel: $^{115}$In NQR line splitting (left scale) due to local internal magnetic field as a function of temperature [(open circles correspond to the splitting $\Delta H_1$ and squares correspond to the splitting $\Delta H_1$ (cf. Fig.~\ref{sp})] and $\mu_{SR}$ frequency (right scale) as a function of temperature (triangles). Lower panel: $\mu_{SR}$ frequency as a function of reduced temperature $(1-T/T_{\rm N})$. The solid lines are the fit to the power-law function Eq.~(\ref{orderparam}).} \label{ord}
\end{figure}

A very intriguing feature of \ICVO\ is the occurrence of two characteristic temperatures in the low-$T$ regime of this compound. We begin the discussion by summarizing how these temperatures manifest in different kinds of measurements. The kink in the static magnetic magnetic susceptibility $\chi(T)$ at 38\,K was originally ascribed to the 3D AFM order of the spins in the bulk of \ICVO\ at $T_{\rm N} = 38$\,K \cite{Kataev2005}, and later on reconsidered to be a signature of local magnetic correlations of the spins at the structural domain boundaries \cite{PhysRevB.78.024420}. Interestingly, the spin-flop transition in the field dependence of the static magnetization at $B_{\rm sf} \sim 5.8$\,T which usually occurs in 3D ordered anisotropic antiferromagnets is observed in \ICVO\ only below 20\,K which has been interpreted as the formation of the 3D N\'eel-type collinear AFM spin structure below this characteristic temperature \cite{PhysRevB.81.060414}. This conclusion was corroborated by the occurrence of the second kink in the static susceptibility and by the observation of the gapped AFM resonance modes at sub-THz frequencies below 20\,K. Such resonance modes are typical for a 3D collinear antiferromagnet, and soften at the same critical field $B_{\rm sf}$ \cite{PhysRevB.81.060414}. However, the gapless paramagnetic ESR signal can be observed only above the upper characteristic temperature of 38\,K \cite{Kataev2005} suggesting that up to this temperature the spin system in \ICVO\ still remains in some quasi-static correlated state featuring anomalous spin dynamics \cite{10.1143/JPSJ.80.023705}. 

Indeed, below the ''upper'' ordering temperature $T_{\rm N} = 39$\,K, as determined by the present local spin probe techniques, the development of the internal field  -- static on the timescale of the NQR and $\mu$SR experiments -- manifests in the splitting of the $^{115}$In NQR lines (Fig.~\ref{sp}) and in the spontaneous precession of the muon spin (Fig.~\ref{ZFspectra}). The splitting of the  $^{115}$In NQR lines enables one to monitor the temperature dependence of the internal field $H_{\rm int}$ probed by the nuclei which is proportional to the sublattice magnetization in \ICVO. 

As to the dynamic characteristics, the $T_1^{-1}$ spin-lattice relaxation rate of $^{115}$In nuclei exhibits a sharp peak at $T_{\rm N} = T^{\ast\ast} = 39$\,K and a broad peak at $T^\ast = 15$\,K (Fig.~\ref{t1nqr}), i.e., it is sensitive to both characteristic temperatures discussed above. The muon relaxation rate exhibits similar characteristic features (Fig.~\ref{ZFlamlong}). In contrast, the $T_1^{-1}$ spin-lattice relaxation rate of $^{51}$V nuclei shows a peak only at $T^\ast = 15$\,K (Fig.~\ref{t1nmr}) due to a special symmetry position of these nuclei (see below).

The internal field $H_{\rm int}$ probed by NQR and $\mu$SR can be considered as a measure of the order parameter of the spin system in \ICVO. Therefore, it is instructive to analyze its temperature dependence in some detail. In NQR, as a quantity proportional to $H_{\rm int}$ we have taken the full splitting of the outer satellites $\Delta H_1$ and $\Delta H_2$ (Fig.~\ref{sp}). Analysis of the $T$-dependence of $\Delta H_i$ can provide  information about the spin dimensionality of the spin system. Such a dependence  is shown  in Fig.~\ref{ord} together with the $T$-dependence of the muon precession frequency which perfectly match together. Both data sets were fitted together by the critical exponent function according to Eq.~(\ref{orderparam}).
The fit yields the ordering temperature $T_{\rm N}$=39.1$\pm$0.1\,K and the critical exponent $\beta$\,=\,0.24$\pm$0.02, practically the same as from the fit of the muon precession frequency alone (cf. Fig.~\ref{ZFfrequency}).   

The theoretically expected values of the critical exponent $\beta$ are $\beta=0.367$ for the 3D Heisenberg spin system, $\beta=0.345$ for the 3D XY-model, $\beta=0.326$ for the 3D Ising system, $\beta=0.231$ for the 2D XY-model, and $\beta=0.125$ for the 2D Ising system  \cite{Jongh1990,Bramwell93,Collins}. The experimentally obtained value of $\beta = 0.24$ is surprisingly close to  the prediction  for the 2D XY model \cite{Bramwell93}, reflecting the predominantly two-dimensional critical behavior of quasi-statically correlated Cu spins in \ICVO. 

It appears from this analysis that, although the occurrence of the internal field may be related to the development of 3D quasi-static correlations -- as it follows from the Quantum Monte-Carlo calculations in Ref.~\cite{PhysRevB.81.060414} -- the correlations develop predominantly in the planes whereas the inter-plane correlation length increases much slower. Furthermore, as it was shown in Ref.~\cite{Bramwell93}, in the framework of the 2D XY model one finds finite magnetization at $T > 0$ in 2D finite-size clusters even if their size approaches a macroscopic scale. This situation is likely to be realized in \ICVO\ since the finite in-plane structural correlation length of $\approx 300$\,\AA\ due to the Cu/V site inversion \cite{PhysRevB.78.024420} may set respective constraints on the magnetic correlation length.
Therefore, one can consider the spin system in \ICVO\ to be in a quasi-2D static state below 39\,K. As it follows from the microscopic model developed in Sect.~\ref{model} such an anisotropic behavior could be attributed to a significant geometrical frustration of the interlayer exchange due to the shift of the Cu-V layers with respect to each other along the $c$-axis. Therefore, the Cu spins in the neighboring layers are AFM coupled on a frustrated triangular motif. In this situation the planes are effectively decoupled but a small amount of defects can partially break the interlayer frustration enabling some correlations also across the planes \cite{Liu2013}. Moreover, as we have shown in Sect.~\ref{model}, in \ICVO\ there is a rather rare situation that the intralayer exchange, being much larger than that between the layers, is not frustrated which stabilizes a quasi-static state below $T^{\ast\ast} = 39$\,K. Due to residual interlayer spin dynamics this is not yet a conventional true 3D N\'eel AFM ordered state. This is evidenced by the pronounced 2D XY critical behavior and also by the absence of the fully developed AFM resonance modes and the field-induced spin-flop transition.  

At first glance, it seems surprising that the slowing down of the in-plane spin dynamics in \ICVO\ by approaching the upper characteristic temperature $T^{\ast\ast} =39$\,K from above does not result in a typical peak in the $T$-dependence of the $^{51}$V relaxation rate $T_1^{-1}$. However, one should keep in mind that the nonmagnetic V ions are located in the plane in the symmetric position with respect to the Cu ions. Therefore, the growth of the in-plane AFM correlations between the Cu spins results in a gradual decrease of the effective local field acting on the V nuclei ultimately down to zero. This yields the slowing down of the  $T_1^{-1}$ rate despite the decreasing frequency of the spin fluctuations. This in-plane dynamics could be probed by V nuclei at the defect non-symmetric crystallographic sites whose amount is however quite small in our samples. Nevertheless the transition into a quasi-2D static state below $T^{\ast\ast}$ is reflected in a reduction of the stretching parameter $b$ from 1 to 0.5, which is typical for a 2D situation (Fig.~\ref{t1nmr}) \cite{Narayanan95}. 
In contrast, a characteristic peak at $T^{\ast\ast}$ is present in the $T_1^{-1}$ rate of $^{115}$In nuclei which are in the non-symmetric position with respect to the Cu spins in the plane (Fig.~\ref{pos}) and the parameter $b$ also drops down to 0.5 at $T^{\ast\ast}$ (Fig.~\ref{t1nqr}). Similarly, the muons at the nonsymmetric interstitial position exhibit at this temperature a sharp peak in the relaxation rate, too (Fig.~\ref{ZFlamlong}).

\begin{figure}[h]
	\includegraphics[width=\columnwidth]{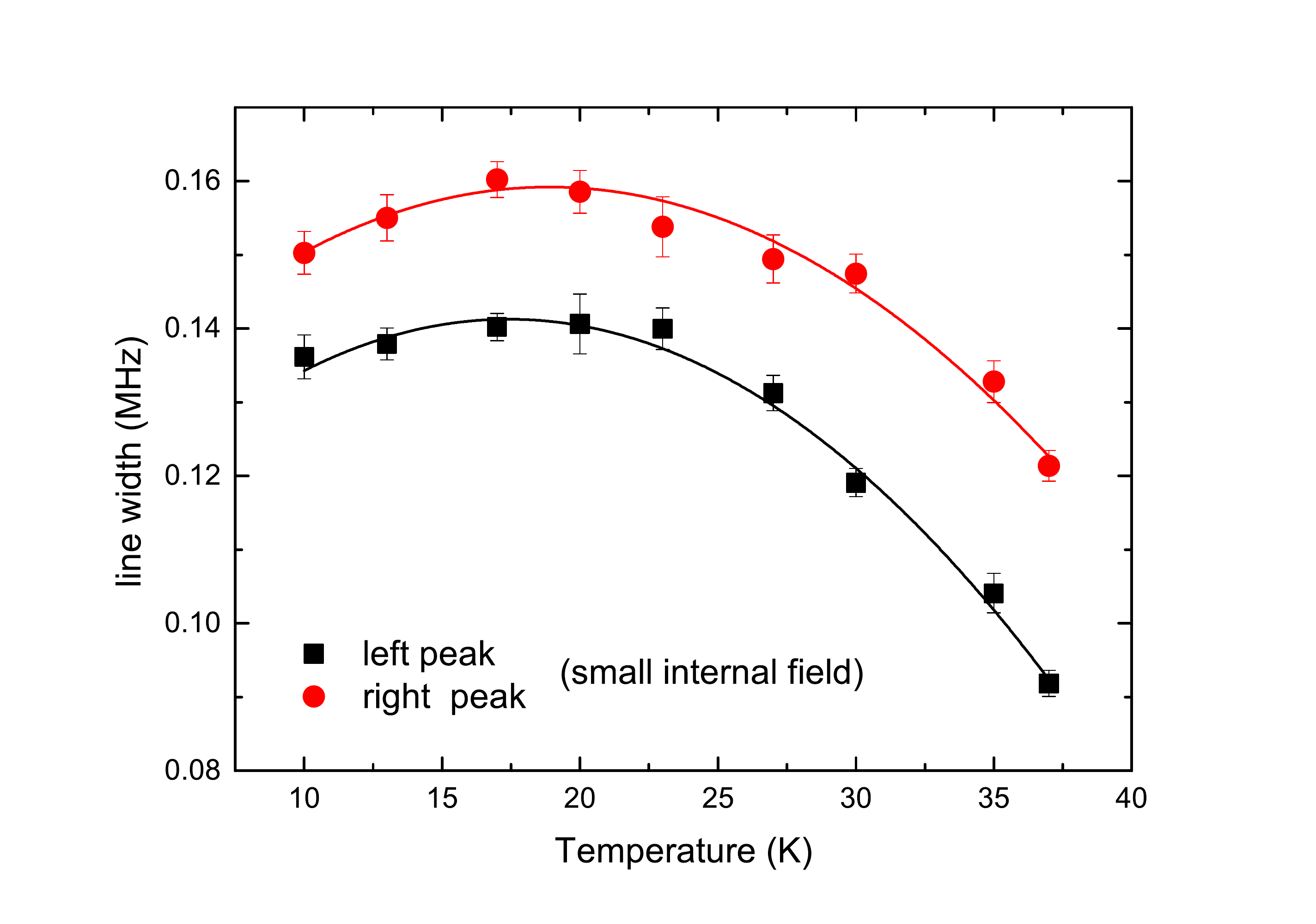}
	\caption{$T$-dependence of the width of the left- and right peaks in the central part of the $^{115}$In NQR spectrum  (Fig.~\ref{HLsp}) corresponding to the In sites exposed to a small internal field. (Solid lines are guides for the eye.)} \label{NQRwidth} 
\end{figure}

The onset of the spin-flop transition below $\sim 20$\,K \cite{PhysRevB.81.060414}  evidences the establishment of the fully developed 3D magnetic order in \ICVO. Due to the shift of the honeycomb planes along the $c$-axis the local fields in the ordered state become nonzero at the V sites as well. Associated with this, the $T$-dependences of the $T_1^{-1}$ rate of both types
of nuclei as well as the relaxation rate of the implanted muons feature the peak at $T^{\ast} = 15$\,K (Figs.~\ref{t1nqr}, \ref{t1nmr} and \ref{ZFlamlong}). This peak is broad suggesting a rather gradual transformation of the magnetic state obviously related to the strong interlayer  frustration and the related residual interlayer spin dynamics which is suppressed only gradually with decreasing temperature. The suppression of this dynamics and the establishment of the 3D order leads to a more homogeneous distribution of local fields. As a consequence, the NQR linewidth which continuously increased below 39\,K saturates and even narrows below $\sim 20$\,K (Fig.~\ref{NQRwidth}).  The absence of an anomaly in the specific heat $C_{\rm p}(T)$ associated with magnetic order  might be related to the fact that a large fraction of magnetic entropy is already lost below 100\,K due to short-range correlations \cite{PhysRevB.78.024420} as well as due to the pronounced two-dimensionality of the spin system which suppresses the $\lambda$-peak in the $C_{\rm p}(T)$ dependence at the magnetic phase transition \cite{Bloembergen77}. Presumably, the ordered moment could be quite small due to the enhanced spin fluctuations in the 2D honeycomb lattice with the low coordination number $z = 3$, which might explain the non-observation of magnetic Bragg peaks in neutron diffraction \cite{PhysRevB.78.024420}. The reason for a missing magnetic anomaly in the thermal expansion \cite{PhysRevB.78.024420} could be a weak magneto-elastic coupling due to the quenched orbital moment of Cu$^{2+}$. 

\begin{figure}[h]
\includegraphics[width=0.9\columnwidth]{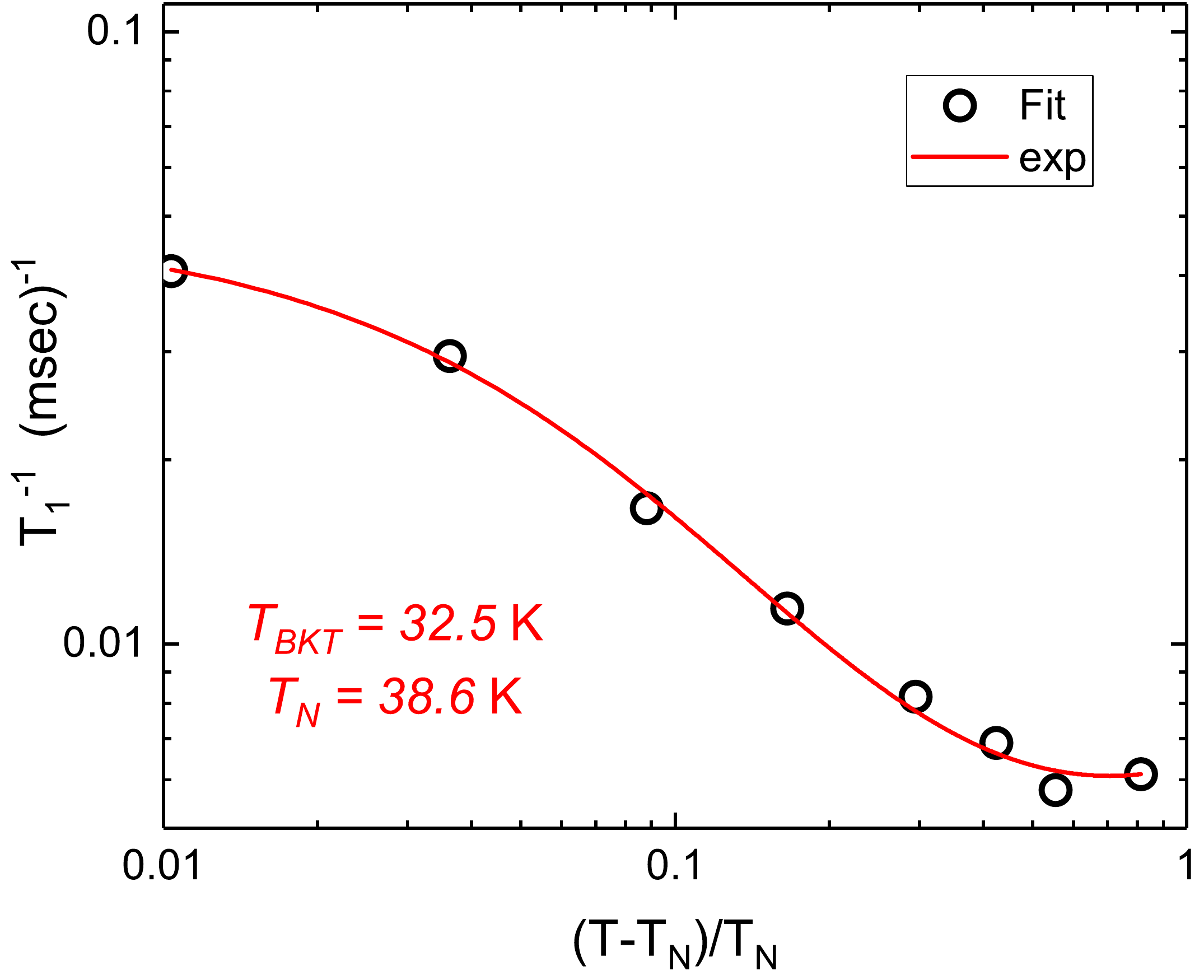}
\caption{$^{115}$In spin-lattice relaxation rate as a function of the reduced temperature $(T-T_{\rm N})/T_{\rm N}$ (cirles) and the fit according to Eq.~(\ref{t1c}) (solid line).} \label{cor}
\end{figure}

The possibility of the realization of the BKT physics in \ICVO\ is certainly a very intriguing issue. That the honeycomb planes in this compound feature planar anisotropy has been shown by our DFT calculations and previously was suggested by ESR experiments in Ref.~\cite{Kataev2005}. It should be noted that for the occurrence of the BKT transition at a finite temperature the pure XY limit is not necessary. As  was shown by Cuccoli {\it et al.} \cite{Cuccoli03} using quantum Monte Carlo simulations, even a very small (of the order of 10$^{-3}$) deviation from the isotropic Heisenberg case towards planar anisotropy  gives rise to the BKT transition at a finite temperature. As has been discussed, e.g., in Ref.~\cite{Regnault1990} the occurrence of the N\'eel order obscures but not necessarily completely excludes the BKT physics. The detection of the BKT transition in low-dimensional quantum spin magnets is a general problem. Unlike in FM or AFM ordered phases magnetization is not the order parameter for the BKT transition. Therefore, the signatures of the BKT physics in such experimental methods as neutron scattering, NMR or ESR can be detected only indirectly by looking for critical exponents above the transition point. The critical behavior of the electron spin system can be studied by analyzing the $T$-dependence of the nuclear spin-lattice relaxation  rate. Having established the 2D XY behavior  below $T_{\rm N}$ = 39\,K  from the analysis of the static internal field in \ICVO,  one can expect that above this temperature the correlation length $\xi(T)_{\rm{BKT}}$  follows an exponential dependence, predicted by Kosterlitz \cite{0022-3719-7-6-005}:
\begin{align}
\xi(T) = \xi_0 {\rm{exp}}(\frac{p}{\sqrt{T/T_{\rm BKT}-1}}),
\label{corl}
\end{align}
where $\xi_0 \sim$ 1  $\AA$ and $p \simeq \pi/2$. $T_{\rm BKT}$ is  the BKT phase transition temperature into the vortex - antivortex  paired state. As it was shown by Borsa $et\, al.$ \cite{PhysRevB.45.5756} 
the spin-lattice relaxation rate in a 2D antiferromagnetic system is proportional to the square of the correlation length $\xi(T)$:

\begin{align}
T_1^{-1} = ^{115}\gamma^2\frac{h_{\rm{eff}}^2}{\omega_e}(\xi(T)/\xi_0)^2.
\label{cl}
\end{align}
Here $^{115}\gamma$ is the gyromagnetic factor of $^{115}$In, $\omega_e$=$\frac{Jk_B}{\hbar}\sqrt{\frac{2zS(S+1)}{3}}$ is the exchange frequency  and  $h_{\rm{eff}}$ is the effective fluctuating hyperfine field \cite{Moriya}. Following the same procedure  as used by Waibel {\it et al.} for the $^{51}$V NMR relaxation in BaNi$_2$V$_2$O$_8$ \cite{PhysRevB.91.214412},  the NQR magnetic spin-lattice relaxation rate for \ICVO\ can be fitted with the following function:
\begin{align}
T_1^{-1} = A [\xi(T)/\xi_0]^2+ kT,
\label{t1c}
\end{align}
where $\xi(T)=\xi(T)_{\rm BKT}$, and the linear $T$-term accounts  for the direct phonon relaxation.   The fit yields $T_{\rm BKT}=32.5 \pm 1.8$\,K  and $p =1.2 \pm0.5$,  which is close to the theoretical value $p\approx 1.6$ within the error bar (Fig.~\ref{cor}).  

Such an estimate of $T_{BKT}$ appears to be consistent with the theoretical results in Ref.~\cite{Bramwell93} where the temperature of the onset of the staggered magnetization (in our case $T_{\rm N}=39$\,K) of a 2D large-size spin cluster is somewhat higher than the BKT transition temperature. 

From the theoretical perspective, our conjecture on the BKT transition in \ICVO\
requires further analysis.  Generally, a BKT transition should manifest
itself in the behavior of long range spin correlations at different
temperatures.  According to our DFT calculations, the minimal model to address
the BKT physics in \ICVO\ is the $S=\frac12$ XXZ model of AB-stacked honeycomb
lattices with the interplane exchange amounting to $\sim$2.5\% of
the nearest neighbor exchange.  Unfortunately, simulating this model is very
challenging: while quantum Monte Carlo techniques suffer from the sign problem
due to magnetic frustration, classical Monte Carlo simulations are not
justified for the extreme quantum case of $S=\frac12$. Nevertheless, we believe
that our experimental indications of the BKT transition in \ICVO\ will
stimulate further numerical studies of this model, for instance, using the
recently introduced pseudofermion functional renormalization group
method~\cite{oj:iqbal16}, capable of treating quantum spin models with
frustration.

\section{Conclusions}\label{conclusion}

To summarize, we have studied the low-temperature magnetic properties of the quasi-2D magnet \ICVO\ featuring  honeycomb planes of AFM coupled Cu$^{2+}$ spins $S = 1/2$ by employing three kinds of local spin probes, the nuclei $^{115}$In and $^{51}$V at the regular lattice sites and  implanted spin-polarized muons $\mu^+$ at interstitial lattice sites. The main objective of this study was to elucidate the nature of the two characteristic  temperature scales of $\sim 40$\,K and $\sim 20$\,K which were controversially interpreted in different previous experiments. The splitting of the $^{115}$In NQR spectral lines and the onset of the muon spin precession at 39\,K evidence the development of the staggered magnetization whose temperature dependence agrees well with the predictions of the 2D XY model suggesting that \ICVO\ is in a quasi-2D static magnetic state below this temperature. A transition to this state is signified by a peak in the $T$-dependence of the relaxation rate $T^{-1}$ of the In nuclei at $T^{\ast\ast} = 39$\,K and a maximum in the longitudinal muon spin relaxation rate, whereas such a peak is absent in the case of the V nuclei due to the cancellation of the local fluctuating fields at this symmetric in-plane position, as expected for the quasi-2D ordered state. However, a true 3D long range magnetic order in \ICVO\ gradually sets in at a significantly lower temperature manifesting in the broad relaxation peaks at $T^\ast = 15$\,K for both types of nuclei and a weak anomaly in the muon relaxation rate. These experimental results are strongly supported by our DFT calculations of the electronic band structure and of the exchange constants in \ICVO\ that reveal the dominance of the nearest neighbor AFM exchange in the honeycomb spin-1/2 planes with a significant XXZ anisotropy and with negligible further neighbor in-plane couplings, as well as a single sizable and frustrated AFM exchange between the honeycomb planes. It appears from our experimental and theoretical findings that such a significant interlayer magnetic frustration concomitant with some structural disorder give rise to two distinct magnetic states successively occurring in \ICVO\ upon lowering the temperature. Particularly intriguing are indications from the analysis of the $^{115}$In relaxation rate $T^{-1}(T)$ of the topological BKT transition in the honeycomb planes of \ICVO\ in the quasi-2D ordered state presumably occurring at $T_{\rm BKT} = 33$\,K, which call for theoretical studies of the BKT physics in this compound. 
  
\section*{Acknowledgments}

This work has been supported in part by the Deutsche Forschungsgemeinschaft in the framework of CRC "Correlated Magnetism - From Frustration to Topology" (SFB~1143, project-id~247310070) and W\"{u}rzburg–Dresden Cluster of Excellence on Complexity and Topology in Quantum Matter – ct.qmat (EXC~2147, project-id~39085490), and is partially based on experiments performed at the Swiss Muon Source S$\mu$S, Paul Scherrer Institute,
Villigen, Switzerland. E.V. and M.I. acknowledge the support of RFBR through grant No. 18-02-00664. O.J. was supported by the Leibniz Association through the Leibniz Competition. A.M. acknowledges support by the Carl Zeiss Foundation. We thank Ulrike Nitzsche for technical assistance.

\section*{Appendix}\label{appendix}

\begin{figure}[b]
	\includegraphics[width=0.9\columnwidth]{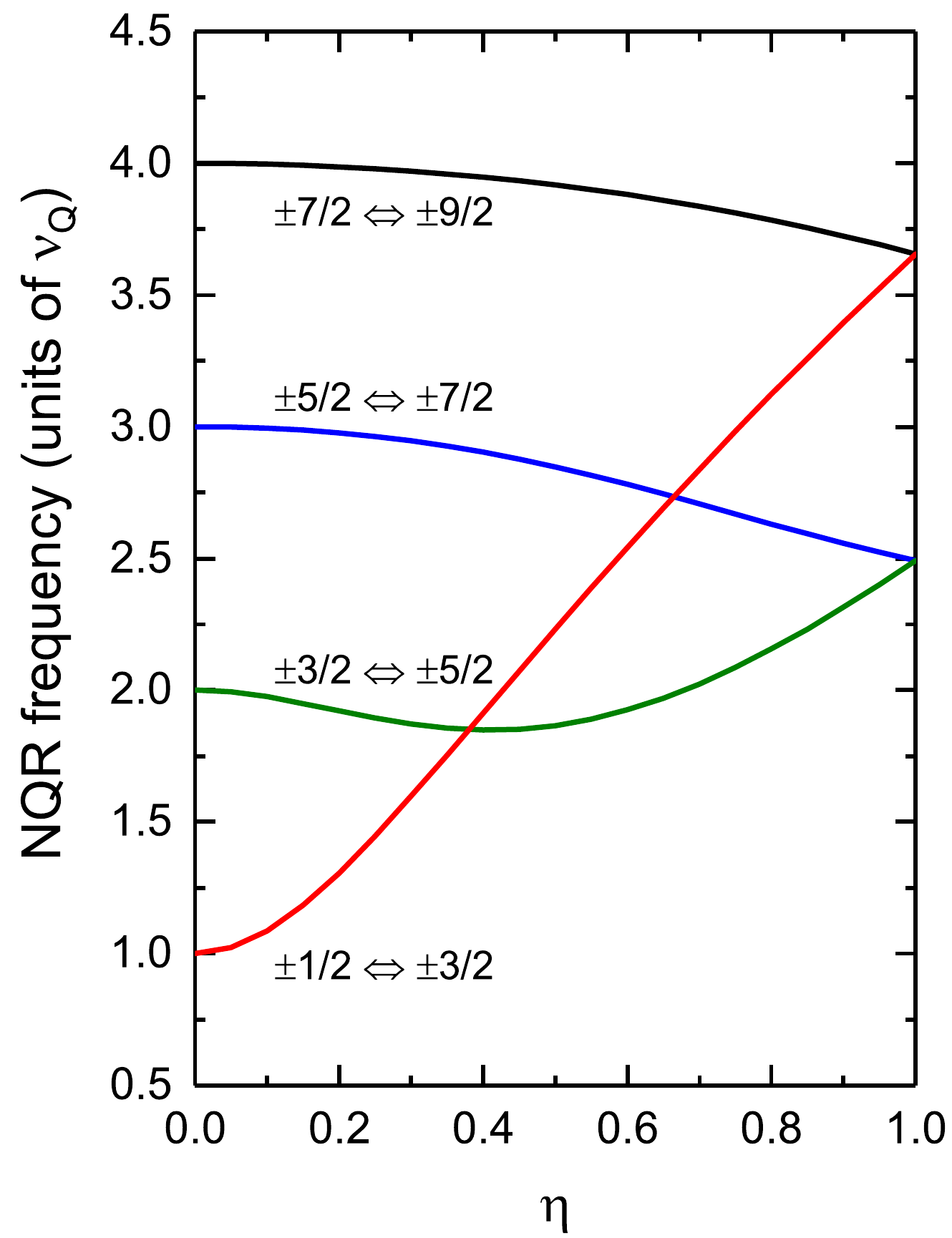}
	\caption{ NQR transition frequencies as a function of the EFG asymmetry parameter $\eta$ numerically calculated using Hamiltonian (\ref{NQRham}).} \label{nu_Q}
\end{figure}
\begin{figure}[b]
	\includegraphics[width=0.9\columnwidth]{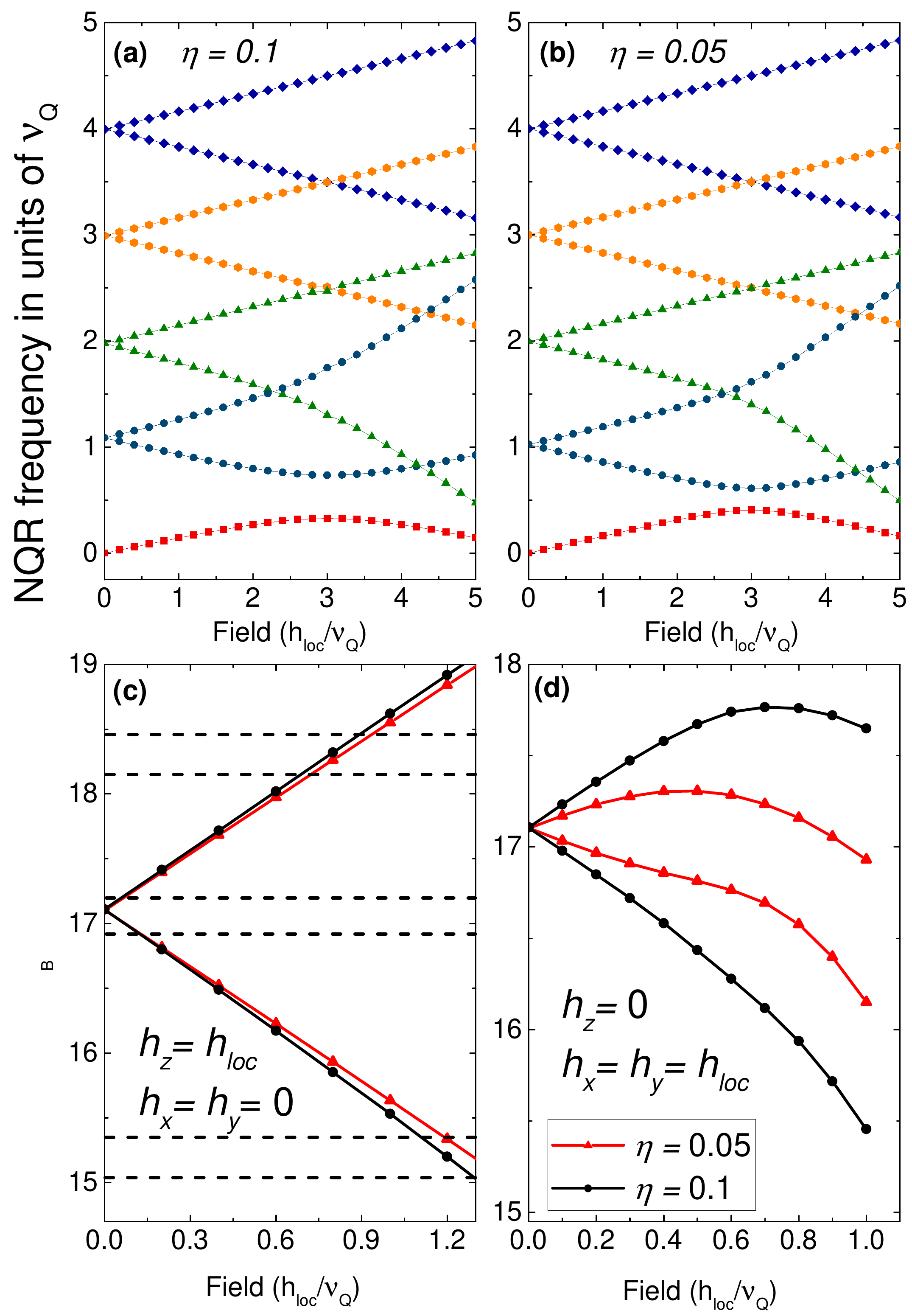}
	\caption{Top: NQR transition frequencies as a function of local magnetic field for the nuclear spin $I$ = $\frac{9}{2}$ in an asymmetric electric field with $\eta = 0.1$ (a) and $\eta = 0.05$ (b). The magnetic field $h$ is applied along the $z$ symmetry axis of EFG. Bottom: (c) Comparison of the calculated frequencies for the $\pm\frac{3}{2}$ $\leftrightarrow$  $\pm\frac{5}{2}$ transition for $h\parallel z$ with the position of the experimentally observed NQR signals at $T=10$\,K indicated by horizontal dashed lines (cf. Fig.~\ref{sp}).  (d) Calculated frequencies for $h\perp z$. Red solid triangles in (c) and (d) correspond to $\eta=0.05$ and black circles correspond to $\eta=0.1$. }     
	
	\label{h_loc}
\end{figure}

Hamiltonian (\ref{NQRham}) can be easily solved for the asymmetry parameter $\eta$ = 0, i.e., in the case of the axial symmetry of the charge distribution around the nucleus. 
For $\eta \not=$ 0 it is still solvable analytically  for the nuclei with $I=3/2$, but for $I>3/2$ there is no analytical solution. Therefore, for  $^{115}$In with $I=9/2$  we have calculated the transition frequencies as a function of $\eta$  numerically.   

The results for the four allowed transitions $\pm1/2\leftrightarrow \pm3/2$, $\pm3/2\leftrightarrow \pm5/2$, $\pm5/2\leftrightarrow \pm7/2$ and $\pm7/2\leftrightarrow \pm9/2$ are presented in Fig.~\ref{nu_Q}. As is customary, the levels are labeled according to the largest component of the wavefunction, though $I_z$ is a good quantum number only when $\eta$ = 0.

Additionally, we numerically solved Hamiltonian (\ref{NQRham}) perturbed by the Zeeman interaction due to the local magnetic field $h$. For calculations we considered two cases of the EFG with $\eta = 0.05$ and $\eta = 0.1$ relevant for \ICVO. Computation results are shown in Figs.~\ref{h_loc}(a) and \ref{h_loc}(b). As a consequence of the Zeeman splitting the additional $-\frac{1}{2}$ $\leftrightarrow$  $\frac{1}{2}$ transition appears. In our calculations we considered the case when the magnetic field is applied along the EFG $z$-principal axis. A comparison of the calculations with the experimental data obtained at 10\,K (cf. Fig.~\ref{sp}) is shown in  Fig.~\ref{h_loc}(c). As it can be seen there, in the case of very small magnetic fields, the line splitting is small and is masked by the overlap of the lines. In stronger fields all four transitions are well separated in a qualitative agreement with experiment. On the quantitative level there is some discrepancy due to the non-collinearity of $h$ and the $z$-axis of EFG which results in an asymmetric splitting of the lines. To illustrate this, in Fig.~\ref{h_loc}(d) we show the other simple limit of $h\perp z$ ($h_x = h_y$).

\bibliography{ICVO_final}

\end{document}